\documentclass[aps,reprint,amsmath, amssymb,groupedaddress,floatfix]{revtex4-1}
\usepackage{natbib}
\usepackage{graphicx}
\usepackage{bm}
\usepackage{color,soul}
\usepackage{hyperref}
\usepackage{float}
\usepackage{csquotes}
\usepackage{color,soul}
\usepackage{hyperref}
\hypersetup{colorlinks=true, urlcolor=blue, citecolor=blue}
\usepackage{longtable}
\usepackage{tabularx}

\begin{document}

\title{First-Principles Property Assessment of Hybrid Formate Perovskites}
\author{Abduljelili Popoola$^1$}
\author{P. S. Ghosh$^{2,3}$}
\author{Maggie Kingsland$^1$}
\author{Ravi Kashikar$^1$}
\author{D. DeTellem$^1$}
\author{Yixuan Xu$^4$}
\author{S. Ma$^4$}
\author{S. Witanachchi$^1$}
\author{S. Lisenkov$^1$}
\author{I. Ponomareva$^1$}
\affiliation{1. Department of Physics, University of South Florida, Tampa, Florida 33620, USA}
\affiliation{2. Glass \& Advanced Materials Division, Bhabha Atomic Research Centre, Mumbai 400 085, India}
\affiliation{3. Homi Bhabha National Institute, Anushaktinagar, Mumbai 400 094, India}
\affiliation{4. Department of Chemistry, University of North Texas, CHEM 305D, 1508 W Mulberry Street, Denton, Texas 76201, USA}
\begin{abstract}
Hybrid organic inorganic formate perovskites, AB(HCOO)$_3$, is a large family of compounds which exhibit variety of phase transitions and diverse properties. Some examples include (anti)ferroelectricity, ferroelasticity, (anti)ferromagnetism, and multiferroism. While many properties of these materials have already been characterized, we are not aware of any study that focuses on comprehensive property assessment of a large number of formate perovskites. Comparison of the materials property within the family is challenging due to systematic errors attributed to different techniques or the lack of data. For example, complete piezoelectric, dielectric and elastic tensors are not available. In this work, we utilize first-principles density functional theory based simulations to overcome these challenges and to report structural, mechanical, dielectric, piezoelectric, and ferroelectric properties for 29 formate perovskites. We find that these materials exhibit elastic stiffness in the range 0.5 to 127.0~GPa , highly anisotropic linear compressibility, including zero and even negative values; dielectric constants in the range 0.1 to 102.1; highly anisotropic piezoelectric response with the longitudinal values in the range 1.18 to 21.12~pC/N, and spontaneous polarizations in the range 0.2 to 7.8~$\mu$C/cm$^2$. Furthermore, we propose and computationally characterize a few formate perovskites, which have not been reported yet.  

\end{abstract}

\maketitle

Hybrid organic inorganic perovskites (HOIP) are receiving a lot of attention presently owing to the rapid progress in their synthesis and characterization. They have the chemical formula ABX$_3$, where A is typically an organic molecule, B is a metallic cation, while X site could be halogen or molecular linker. They exhibit a variety of phase transitions and rich range of properties, such as ferromagnetism, (diel/ferro)ectricity, non-linear optical properties, caloric effects, ferroelasticity, multiferroicity, among others \cite{Stroppabook, MOF_mag}. Among chemically diverse HOIPs, AB(HCOO)$_3$, is one of the largest families, where one can find most of the aforementioned properties. The structural phase transitions in these materials are primarily driven by the hydrogen bond stabilization and often occur close or even above room temperature, which is highly desirable feature \cite{Stroppabook}. For example, most ethyl ammonium metal formate perovskites exhibit transition in the range 293 to 400~K \cite{10.1002/chem.201303425, https://doi.org/10.1002/anie.201510024, C5DT04536C, C6CP05151K}. Magnetic properties of formates are mostly determined by weak magnetic interactions mediated by formate linker causing them to exhibit magnetic ordering at low temperatures only, typically below 50~K \cite{Stroppabook}. Furthermore, [AZE][M(HCOO)$_3$] (AZE = azetidinium; M = Mn$^{2+}$, Cu$^{2+}$ and Zn$^{2+}$) family was reported to have extraordinarily large dielectric constants higher than 10$^4$ in the vicinity of room temperature \cite{ref51Stroppabook,ref53Stroppabook,ref54Stroppabook}. Often times, the value exhibit strong frequency dependence, which resemble behavior of ferroelectric relaxors  \cite{https://doi.org/10.1111/j.1551-2916.2011.04952.x}.    
Many formates undergo transitions into polar space groups and, therefore, are possible candidates for ferroelectricity, which is defined by the presence of spontaneous electric polarization reversible by electric field. However, the value of spontaneous polarization is typically below 5~$\mu$C/cm$^2$, which makes its experimental measurement very challenging \cite{P4}. Table \ref{tab:exp-pol1} provides polarization values from the literature and conditions for which it was reported/computed.
The simultaneous realization of ferroelectricity, ferromagnetism and/or ferroelasticity in some hybrid formates classify them as multiferroics. It was shown that DMA-Zn(HCOO)$_3$ becomes multiferroic on substitution of Zn with transition metals such as Ni, Mn, Co and Fe \cite{ref32Stroppabook,ref42Stroppabook, MOF_mag}. DMA-Co(HCOO)$_3$ is another hybrid in which multiferroicity has been observed \cite{ref48Stroppabook}. All abbreviations for A sites used in this study are listed in Table \ref{tab:abbreviations}.

  \begin{table*}
    \centering
    \caption{Abbreviations for A sites used in the study, following Stroppas book\cite{Stroppabook}}
    \begin{ruledtabular}
    \begin{tabular}{cccccccc}
         Material & NH$_2$NH$_3$ & C$_2$H$_5$NH$_3$ & C(NH$_2$)$_3$ & (CH$_3$)$_2$NH$_2$ & CH$_3$NH$_3$ & NH$_2$CHNH$_2$\\
         \hline
         abbreviation & HAZ & EA & Gua & DMA & MA & FA\\
    \end{tabular}
    \end{ruledtabular}
    \label{tab:abbreviations}
\end{table*}

  \begin{table*}
    \centering
    \caption{Polarizations from literature and those computed in this work. A = NH$_3$CH$_2$CH$_3$, P = PH$_3$CH$_2$CH$_3$, AF = NH$_3$CH$_2$CF$_3$, PF = PH$_3$CH$_2$CF$_3$.}
    \begin{ruledtabular}
    \begin{tabular}{cccccc}
         Material & Type & Literature & Conditions & P ($\mu$C/cm$^2$) \\
         \hline
         A-Mn & Comp.& 1.80$^{\text{\cite{ref84Stroppabook}}}$& T = 0~K & -- \\
         P-Mn &Comp.& 1.00$^{\text{\cite{ref84Stroppabook}}}$ & T = 0~K & --\\  
         AF-Mn &Comp.& 5.10$^{\text{\cite{ref84Stroppabook}}}$ & T = 0~K & --\\ 
         PF-Mn &Comp.& 5.90$^{\text{\cite{ref84Stroppabook}}}$ & T = 0~K & --\\
         \hline
         DMA-Mn &Exp.& 0.30$^{\text{\cite{C8RA00799C}}}$ & T = 150~K; B = 9~T (during growth) & 7.52 \\
          &Exp.& 2.70 -- 3.61$^{\text{\cite{WangW2013Mcit, doi:10.1021/acs.jpcc.0c03916}}}$ & T = 150~K; B = 0 -- 5~T; E = 5~kV/cm &  -- \\
          &Exp.& 0.8-2.4$^{\text{\cite{doi:10.1063/1.4989783}}}$ & T = 184~K; E = 3.1-7.7~kV/cm & -- \\
         DMA-Ni &Exp& 0.42 -- 0.52$^{\text{\cite{Ma_2019}}}$ & T = 150~K; B = 0 -- 10~T & -- \\
         DMA-Co &Exp.& 0.30$^{\text{\cite{doi:10.1021/acs.jpclett.0c02943}}}$ & T = 125~K & 7.44 \\
         DMA-Zn &Exp.& 0.45$^{\text{\cite{C6CP03414D}}}$  & T = 125~K & 7.77 \\
         \hline
         NH$_4$Mn &Exp.& 0.97$^{\text{\cite{doi:10.1021/ja206891q}}}$ & T = 140 ~K & 2.45   \\
         NH$_4$Mg &Exp.& 1.15$^{\text{\cite{10.1002/chem.201303425}}}$ & T = 93~K  & --  \\
         NH$_4$Zn &Exp.&  0.02 -- 0.93$^{\text{\cite{doi:10.1021/ic4020702}}}$ & T = 120 -- 248~K & 2.43 \\
         &Exp.& 4.00$^{\text{\cite{C6CE01891B}}}$ & T = 273~K; P = 1.44~GPa & -- \\
         &Exp.& 1.03$^{\text{\cite{doi:10.1021/ja104263m, doi:10.1021/ja206891q}}}$ & T = 163~K & -- \\
         NH$_4$Sc  &Comp.& 3.71$^{\text{\cite{C6RA24182D}}}$ & T = 0~K & -- \\
         NH$_4$Ti  &Comp.& 2.46$^{\text{\cite{C6RA24182D}}}$ & T = 0~K & -- \\
         NH$_4$V  &Comp.& 2.40$^{\text{\cite{C6RA24182D}}}$ & T = 0~K & --\\
         NH$_4$Cr  &Comp.& 2.51$^{\text{\cite{C6RA24182D}}}$ & T = 0~K & --\\
         NH$_4$Mn  &Comp.& 2.38$^{\text{\cite{C6RA24182D}}}$ & T = 0~K & --\\
         NH$_4$Fe  &Comp.& 2.37$^{\text{\cite{C6RA24182D}}}$ & T = 0~K& --\\
         NH$_4$Co  &Comp.& 2.36$^{\text{\cite{C6RA24182D}}}$ & T = 0~K&--\\
         NH$_4$Ni  &Comp.& 2.17$^{\text{\cite{C6RA24182D}}}$ &T = 0~K& 2.27\\
         NH$_4$Cu  &Comp.& 2.20$^{\text{\cite{C6RA24182D}}}$ & T = 0~K&--\\
         NH$_4$Zn  &Comp.& 2.30$^{\text{\cite{C6RA24182D}}}$ &T = 0~K &2.43\\
         \hline
         CH$_3$NH$_2$NH$_2$Mn&Exp.& 0.14$^{\text{\cite{doi:10.1021/acs.chemmater.6b05249}}}$ & T = 150~K; B = 10~T & -- \\
         \hline
         NH$_3$(CH$_2$)$_4$NH$_3$Mg$_2$ &Exp.& 1.51$^{\text{\cite{10.1002/chem.201303425}}}$ & T = 93~K & -- \\
         \hline
         HAZ-Mn &Exp. (estimated)& 3.58$^{\text{\cite{C3QI00034F}}}$ & T = 110~K & 2.17\\
         HAZ-Co &Exp. (estimated)& 2.61$^{\text{\cite{C3QI00034F}}}$ & T = 405~K & 2.53\\
         HAZ-Zn &Exp. (estimated)& 3.48$^{\text{\cite{C3QI00034F}}}$ & T = 110~K & 2.32\\
         HAZ-Mg &Exp. (estimated)& 3.44$^{\text{\cite{C3QI00034F}}}$ & T = 400~K & 2.53\\
         &Comp.& 2.6$^{\text{\cite{doi:10.1021/acs.jpcc.6b10714}}}$ & T = 150 -- 375~K & 2.53 \\
         \hline 
         EA-Mg &Exp. (estimated) & 3.43$^{\text{\cite{10.1002/chem.201303425}}}$ & T = 93~K & 1.26 \\
         \hline
         Gua-Cr &Comp.& 0.22$^{\text{\cite{StroppaA2013HIFi}}}$ & T = 0~K&-- \\
         Gua-Cu$_{0.5}$Mn$_{0.5}$ &Comp.& 9.90$^{\text{\cite{PhysRevMaterials.2.014407}}}$&T = 0~K &-- \\
         Gua-Cu&Comp.& 0.11 - 0.37$^{\text{\cite{2015PSSRR962T, ref79Stroppabook}}}$ &T = 0~K & 0.21\\
    \end{tabular}
    \end{ruledtabular}
    \label{tab:exp-pol1}
\end{table*}

Mechanical properties have been investigated for several members of formate families and are reviewed in Ref.\cite{ref101Stroppabook}. Some representative data from the literature for Young's and elastic moduli are compiled in Table \ref{tab:Exp_elastics}. The exotic negative linear compressibility, defined as an increase in lattice parameter(s) under hydrostatic pressure, has been computationally predicted in HAZ-M(HCOO)$_3$ (M = Mn,Fe,Co) \cite{ghosh2021,ghosh2022} and NH$_4$Zn(HCOO)$_3$ \cite{doi:10.1021/ja305196u}. Negative linear compressibility finds applications in pressure sensors and actuators, and possibly in design of artificial muscles\cite{NLC1}.
\begin{table*}
    \centering
    \caption{Elastic properties from the literature. AZE = (CH$_2$)$_3$NH$_2$}
    \begin{ruledtabular}
    \begin{tabular}{ccccc}
        Material & Type & Young's Moduli (GPa) & Elastic Moduli (GPa) & Ref. \\
        \hline
        DMA-Ni & Exp. & 24.5 & & \cite{ref101Stroppabook} \\
        DMA-Mn & Exp. & 19.0 & & \cite{ref101Stroppabook} \\
        DMA-Co & Exp. & 21.5 & & \cite{ref101Stroppabook} \\
        DMA-Zn & Exp. & 19.0 & & \cite{ref101Stroppabook} \\
        \hline
        Gua-Cu & Exp. & & 15.0 -- 21.0 & \cite{ref103Stroppabook} \\
        Gua-Zn & Exp. & & 24.0 -- 29.0 & \cite{ref103Stroppabook} \\
        Gua-Mn & Exp. \& Comp. & & 23.5(6) -- 28.6(4) & \cite{ref106Stroppabook} \\
        \hline
        AZE-Cu & Exp. \& Comp. & & 11.5(4) -- 12.6(3) & \cite{ref106Stroppabook} \\ 
        \hline
        HAZ-Zn & Exp. & & 24.5 -- 26.5 & \cite{doi:10.1021/acs.chemmater.5b04143} \\
        HAZ-Mn & Exp. & & 24.5 -- 28.6 & \cite{doi:10.1021/acs.chemmater.5b04143} \\
        \hline
        NH$_4$Zn & Exp. \& Comp. & & 18.2 -- 34.4 & \cite{doi:10.1021/ja305196u} \\
    \end{tabular}
    \end{ruledtabular}
    \label{tab:Exp_elastics}
\end{table*}

Evidences of pyroelectricity have reported in some hybrid formates perovskites. For instance, pyroelectric coefficient was measured to exhibit a maximum of 5.16$\times$10$^{-2}$~C/m$^2$~K under a poling electric field of 7.7~kV/cm at 192~K for DMA-Mn(HCOO)$_3$ \cite{doi:10.1063/1.4989783}.
In another instance, pyroelectric current was reported and used to study the order-disorder transition under different pressures in DMA-Co(HCOO)$_3$ \cite{doi:10.1021/acs.jpclett.0c02943}. Some other hybrid formates perovskites in which pyroelectric current has been measured include DMA-Mg(HCOO)$_3$ \cite{doi:10.1021/acs.jpcc.0c04505}, DMA-Mn(HCOO)$_3$, DMA-Mn$_{0.5}$Ni$_{0.5}$(HCOO)$_3$ \cite{doi:10.1021/acs.jpcc.0c03916, WangW2013Mcit}, Gua-Cu(HCOO)$_3$ \cite{2015PSSRR962T}, CH$_3$NH$_2$NH$_2$Mn(HCOO)$_3$ \cite{doi:10.1021/acs.chemmater.6b05249} and DMA-Zn(HCOO)$_3$ \cite{C6CP03414D}. The dependence of pyroelectric current on applied magnetic field has also been demonstrated in DMA-Ni(HCOO)$_3$ \cite{Ma_2019}.

Although, the aforementioned studies highlight the outstanding progress that has been made in characterization of these materials, the survey also reveals scarcity of such investigations, especially in the light of the fact that formates subgroup hosts at least 64 known members \cite{Stroppabook,P5,P6}. It should also be recognized that many such characterizations, spontaneous polarization for example, are rather challenging experimentally. On the other hand, computational investigation is an inexpensive, reliable and efficient tool to overcome these challenges and achieve a comprehensive assessment of structural, piezoelectric, dielectric and elastic properties for a wide range of materials in the formate family. 

Therefore, in this study, we aim: (i) to predict structural parameters, polarization, piezoelectric coefficients, dielectric constants and elastic stiffness of 29 formate compounds using first-principles density functional theory (DFT) based simulations; (ii) to provide a comprehensive comparative assessment of the aforementioned properties; (iii) to catalog the properties which could aid screening of promising materials. 

\section{Computational Methodology}
\begin{table*}

    \caption{Materials investigated in this study (upper part of the Table). The underscored compounds do not have experimental structure available from the literature. The bottom part of the Table gives other materials from the same family, which are not part of this investigation.}
    \begin{ruledtabular}
    \begin{tabular}{cccc}
     Material & Space Group& T (K) & Experimental Ref. \\
        Gua-M (M=Co, Fe, Mn, Ni, Zn) & Pnna& 293 &  \cite{10.1002/chem.200901605} \\
        Gua-Cu & Pna2$_1$& 293 & \cite{10.1002/chem.200901605} \\
        
        EA-Mg & Pna2$_1$& 292 & \cite{10.1002/chem.201303425} \\
        
        DMA-M (M = \underline{Co}, Mn, \underline{Zn}) & Cc & 285 & \cite{DMA-Mn} \\
        MA-M (M = Mn, Zn, Co, Ni) & Pnma & 135, 290, 290, 100 & \cite{10.1002/asia.201200139, 10.1002/chem.201703140, C6TC03992H} \\
        HONH$_3$M (M=Co, Mg, Mn, Ni, Zn, \underline{Fe}) & P2$_1$2$_1$2$_1$ & 293 & \cite{doi:10.1021/ic302129m} \\
        FA-Mn & C2/c & 110 &\cite{doi:10.1021/ic500479e} \\
        HAZ-M (M = Mn, Zn, Co)  & Pna2$_1$ & 110, 110, 298 & \cite{C3QI00034F, C7DT02927F} \\
        HAZ-M (M = Mg)  & P2$_1$2$_1$2$_1$ & 110 & \cite{C3QI00034F} \\
         NH$_4$M (M=Co, Fe, Mn, Zn) & P6$_3$ & 110 & \cite{doi:10.1021/ja206891q} \\
    \hline
        Gua-M (M = Cd) & R$\overline{3}$c & 150 & \cite{Gua-Cd}\\
        Gua-M (M = Zn) & Pnna & 120 & \cite{Gua-Zn}\\
        NH$_4$M (M = Cu) & P2$_1$2$_1$2$_1$  & 120 & \cite{NH4-Cu}\\
        NH$_4$M (M = Mg) & P6$_3$  & 93 & \cite{NH4-Mg}\\
        
    \end{tabular}
    \end{ruledtabular}
    \label{tab:exp}
\end{table*}
 
Table \ref{tab:exp} lists the hybrid formates that we have investigated and the associated experimental references from which the structures have been retrieved along with temperatures at which the structures were recorded. The bottom part of the panel lists some of the HOIPs in the same family, which did not become part of this study. The experimental structures were used to initialize DFT based computations as implemented in VASP package \cite{VASP1,VASP2,VASP3,VASP4}. Technically, all experimental structures were first fully relaxed using Perdew-Burke-Ernzerhof (PBE) version of the generalized gradient approximation for exchange correlation functional \cite{PBE}. In order to model hydrogen bonds, we used dispersion corrections of zero-damping D3 \cite{grimme06, grimme10}, which was previously shown to provide good agreement with experimental structures \cite{doi:10.1021/acs.jpcc.1c03980,PhysRevLett.128.077601,PhysRevLett.125.207601,doi:10.1021/acs.jpcc.1c01138,PhysRevB.104.235132}. The electron-ion interactions are treated with the projected augmented wave (PAW) potentials \cite{blochl94}. We used the plane wave cutoff energy in range 700-850~eV and non-Gamma centered k-point mesh which corresponds to k-point densities in range 0.19 -- 0.57~\AA$^{-1}$. 
Note that k-point density and cutoff energy for each material are given in Table S1 in supplementary material. 
The Hubbard correction as proposed by Dudarev {\it et al.} \cite{PhysRevB.57.1505} is introduced to account for the Coulomb repulsion between localized d-electrons of transition metals. Unit cell parameters and atomic positions were relaxed until stress and forces are less than 0.1~GPa and 1~meV/$\AA$, respectively.  The energy convergence criterion for self-consistent calculations was 10$^{-6}$~eV.
The crystal polarization is evaluated by the Berry phase method developed by King-Smith and Vanderbilt \cite{PhysRevB.47.1651,PhysRevB.48.4442}. We computed the {\it intrinsic} piezoelectric constants $e_{ij}$ and $d_{ij}$ (in matrix notations) defined as the linear response of the polarization to the applied strain and stress, respectively. The $d_{ij}$ coefficients were obtained from  $d_{ij}=e_{ik}(C^{-1})_{kj}$, where $C$ is the single crystal elastic constant matrix. The constants  $e_{ik}$ and $C_{ij}$ were computed using finite difference method as implemented in VASP \cite{PhysRevB.65.104104}. Hubbard U were employed for transition metal atoms. The following values computed using the linear response ansatz of Cococcioni et al \cite{dftu} from PAW approach in VASP were utilized:  6.5~eV for Mn, 7.2~eV for Fe , 4.6~eV for Co and 5.1~eV for Ni.

\section{Results and Discussion}

\begin{table*}
    \centering
    \footnotesize
    \caption{Structural properties and polarizations.  Ground state magnetic ordering of magnetic materials is in parenthesis, where FM and AFM denote ferromagnetic and antiferromagnetic orderings, respectively.  The data in paranthesis give experimental values taking from the reference given in the last column.}
    \begin{ruledtabular} 
    \resizebox{0.85\textwidth}{!}{\begin{tabular}{ccccccccccccccc}
         material & space group & a(\AA) & b(\AA)  & c(\AA)  & $\beta$ ($^{o}$) &  V (\AA$^3$) & $\mu_B$ & \textbf{P} ($\mu$C/cm$^2$)  & ref  \\
         \hline
         Gua-Mn (AFM) & Pnna  & 8.42 (8.52) & 11.91 (11.98) & 9.12 (9.06) & 90 & 915 (925)&4.7 & non-polar &\cite{10.1002/chem.200901605} \\
         Gua-Fe (AFM) & Pnna  & 8.35 (8.42) & 11.78 (11.85) & 8.98 (8.95) & 90 & 883 (892)&3.8 & non-polar&\cite{10.1002/chem.200901605}\\
         Gua-Co (AFM) & Pnna & 8.28 (8.33) & 11.64 (11.75) & 8.99 (8.91) & 90 & 866 (873)&2.8 & non-polar& \cite{10.1002/chem.200901605}\\
         Gua-Ni (AFM) & Pnna  & 8.24 (8.26) & 11.63 (11.64) & 8.91 (8.83) & 90 & 853 (850)& 1.8 & non-polar&\cite{10.1002/chem.200901605} \\
         Gua-Cu (AFM) & Pna2$_1$  & 8.50 (8.52) & 9.07 (9.03) & 11.27 (11.35) & 90 & 869 (874)&0.6 & (0, 0, 0.21) (0.11--0.37)&\cite{10.1002/chem.200901605, ref79Stroppabook,2015PSSRR962T}\\
         Gua-Zn & Pnna  & 8.27 (8.35) & 11.66 (11.73) & 8.99 (8.91) & 90 & 868 (872)&0.0 & non-polar&\cite{10.1002/chem.200901605} \\
         \hline
         HONH$_3$-Mn (AFM) & P2$_1$2$_1$2$_1$  & 7.80 (7.81) & 8.04 (7.96) & 13.06 (13.17) & 90 & 819 (819)&4.5 & non-polar&\cite{doi:10.1021/ic302129m} \\
         HONH$_3$-Co (AFM) & P2$_1$2$_1$2$_1$  & 7.67 (7.68) & 7.82 (7.76) & 13.00 (13.02) & 90 & 780 (776)&2.7 & non-polar&\cite{doi:10.1021/ic302129m} \\
         HONH$_3$-Ni (AFM) &  P2$_1$2$_1$2$_1$  & 7.59 (7.62) & 7.98 (7.78) & 12.80 (12.73) & 90 & 775 (755)&1.8 & non-polar&\cite{doi:10.1021/ic302129m} \\
         \underline{HONH$_3$-Fe (AFM)} &  P2$_1$2$_1$2$_1$ & 7.70 & 8.00 & 13.05 & 90 & 802 & 3.6 & non-polar \\
         HONH$_3$-Zn & P2$_1$2$_1$2$_1$  & 7.65 (7.69) & 7.83 (7.74) & 13.18 (13.02) & 90 & 790 (775)&0.0 & non-polar&\cite{doi:10.1021/ic302129m} \\
         HONH$_3$-Mg &P2$_1$2$_1$2$_1$  & 7.67 (7.69) & 7.89 (7.79) & 12.73 (12.86) & 90 & 770 (770)&0.0 & non-polar&\cite{doi:10.1021/ic302129m} \\
         \hline
         HAZ-Co (AFM) & Pna2$_1$  & 8.63 (8.65) & 7.74 (7.76) & 11.46 (11.55) & 90 & 765 (776)&2.7 & (0, 0, 2.81) (2.61 at 405~K)&\cite{C7DT02927F} \\ 
         HAZ-Mn (FM) & Pna2$_1$ & 8.99 (8.93) & 7.83 (7.82) & 11.66 (11.69) & 90 & 820 (817) & 4.7& (0, 0, 2.51) (3.58 at 110~K)&\cite{C3QI00034F}\\
         HAZ-Zn & Pna2$_1$  & 8.65 (8.66) & 7.75 (7.72) & 11.49 (11.48) & 90 & 771 (768) &0.0& (0, 0, 2.59) (2.6--3.48 at 0--110~K) &\cite{C3QI00034F, doi:10.1021/acs.jpcc.6b10714}\\
         \underline{HAZ-Mg}& Pna2$_1$  & 8.87  & 7.63  & 11.45  & 90 & 775 &0.0 & (0, 0, 2.90) (3.44 at 400~K)&\cite{C3QI00034F}\\
         HAZ-Mg & P2$_1$2$_1$2$_1$  & 8.00 (7.89) & 13.90 (13.75) & 7.28 (7.38) & 90 & 809 (802)&0.0 & non-polar&\cite{C3QI00034F} \\
         \hline
         NH$_4$-Co (AFM) & P$_3$  & 12.54 (12.59) & 12.54(12.59) & 8.34 (8.22) & 90 & 1136 (1128)&2.6 & (0, 0, 2.35)&\cite{doi:10.1021/ja206891q} \\
         NH$_4$-Fe (AFM)& P6$_3$  & 12.58 (12.62) & 12.58 (12.62) & 8.57 (8.36) & 90 & 1174 (1153)& 3.8 & (0, 0, 2.41)&\cite{doi:10.1021/ja206891q} \\
         NH$_4$-Mn (AFM)& P6$_3$  & 12.55 (12.67) &  12.55 (12.67)  & 8.71 (8.54) & 90 &  1189 (1187)&4.5 & (0, 0, 2.45)&\cite{doi:10.1021/ja206891q}\\
         NH$_4$-Zn& P6$_3$  & 12.56 (12.59) & 12.56 (12.59) & 8.38 (8.20) & 90 & 1144 (1126)&0.0 & (0, 0, 2.43) (0.93--1.03 at 120--163~K)&\cite{doi:10.1021/ja206891q, doi:10.1021/ja104263m, doi:10.1021/ja206891q, doi:10.1021/ic4020702} \\
         \hline
         MA-Co (AFM) & Pnma  & 8.18 (8.28) &11.67 (11.67) & 8.28 (8.15) & 90 & 790 (789)&2.8 & non-polar&\cite{10.1002/chem.201703140} \\
         MA-Co (AFM) & P2$_1$/c  & 8.25 (8.18) &11.69 (11.67) & 8.20 (8.15) & 93.6 (91.9) & 789 (789)&2.8 & non-polar&\cite{10.1002/chem.201703140} \\
         MA-Mn (AFM) & Pnma  & 8.39 (8.68) & 11.93 (11.95) & 8.42 (8.17) & 90 & 843 (847)&4.7 & non-polar&\cite{10.1002/asia.201200139}\\
         \underline{MA-Mn} (AFM) & P2$_1$/c  & 8.41 & 8.47 & 14.12 & 121.9 & 853 &4.7 & non-polar&\\     
         MA-Ni (AFM) & Pnma  & 8.15 (8.18) &11.62 (11.52) & 8.24 (8.08) & 90 & 781 (762)&1.8  & non-polar&\cite{C6TC03992H} \\
         MA-Zn & Pnma  & 8.31 (8.41) & 11.69(11.71) & 8.17 (8.10) & 90 & 794 (798)&0.0 & non-polar&\cite{10.1002/asia.201200139} \\
         \underline{MA-Zn} & P2$_1$/c  & 8.27 & 13.89 & 8.26 & 122.4 & 802 &0.0 & non-polar& \\
         \hline
         DMA-Co (AFM) &   Cc & 14.10  & 8.24  & 8.62  & 121 & 859 &2.7& (6.93, 0, 2.70)(0.3 at 125~K)&\cite{doi:10.1021/acs.jpclett.0c02943}\\
         DMA-Mn (AFM) & Cc & 14.35 (14.35) & 8.39 (8.32) & 8.86 (8.88) & 120.3 & 922 (912)&4.7 & (7.10, 0, 2.48) (0.3--2.7 at 150--184~K)&\cite{DMA-Mn, doi:10.1021/acs.jpcc.0c03916, doi:10.1063/1.4989783, WangW2013Mcit, C8RA00799C} \\
         DMA-Zn & Cc & 14.13 & 8.24 & 8.64 &120.71 & 865 &0.0 & (7.30, 0, 2.66) (0.45 at 125~K)&\cite{C6CP03414D}\\
         \hline
         EA-Mg & Pna2$_1$ & 8.82 (8.90) & 8.12 (8.12) & 11.73 (11.81) & 90 & 840 (853)&0.0& (0, 0, 1.45) (3.34 at 93~K)&\cite{10.1002/chem.201303425} \\
         \hline
         FA-Mn (AFM) &C2/c & 13.12(13.44) & 8.86(8.68) & 8.23(8.41) & 118.1 & 859(865)& 4.7 & non-polar&\cite{doi:10.1021/ic500479e} \\
    \end{tabular}}
    \end{ruledtabular}
    \label{table:pol1}
\end{table*}
\subsection{Structure}
\begin{figure*}
    \centering
    \includegraphics[width=0.8\textwidth]{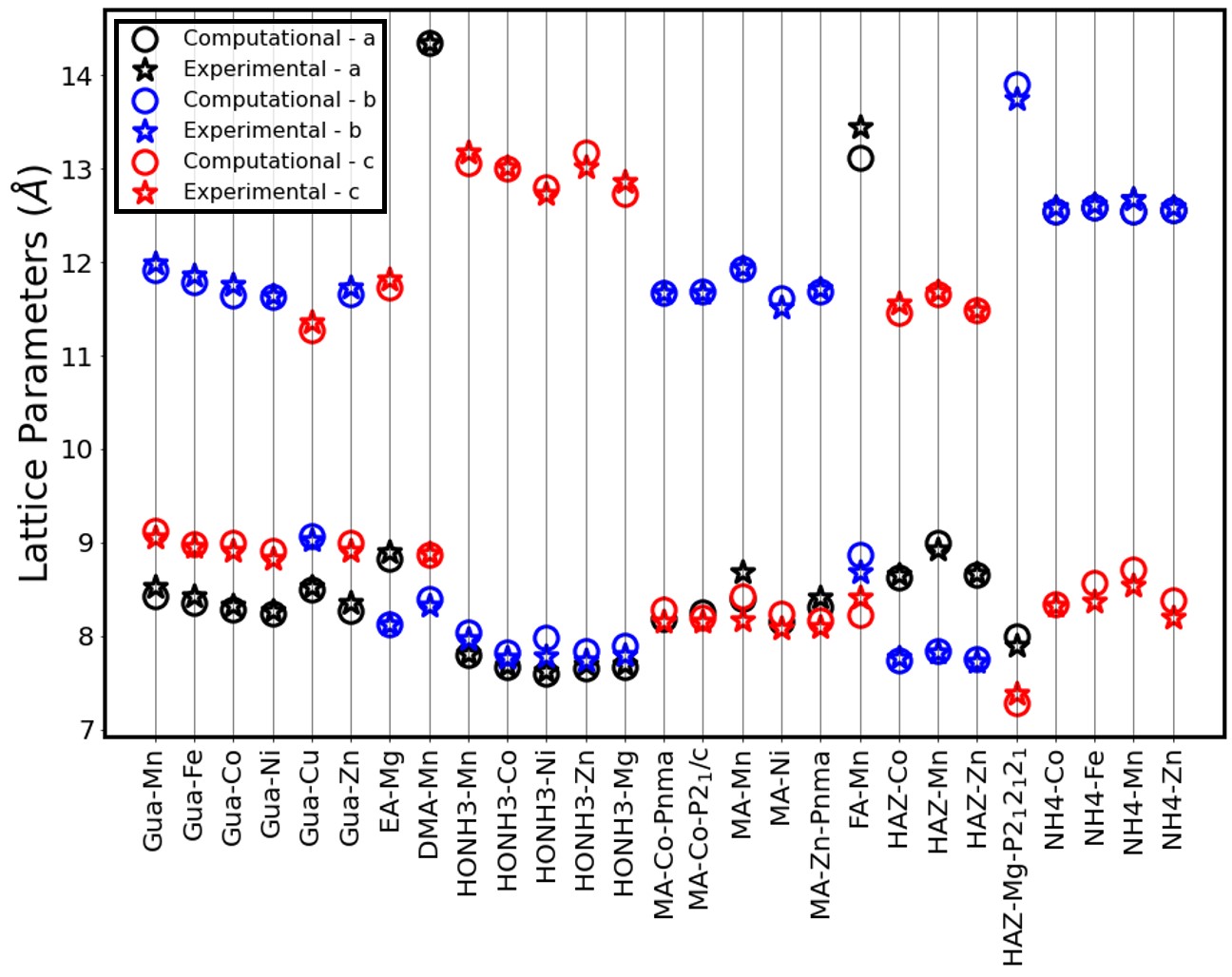}
    \caption{Comparison between computational and experimental lattice parameters.  Only the structures where space group is the same for both computations and experiment are compared. }
    \label{fig1}
\end{figure*}

The ground state structural parameters are reported in Table {\ref{table:pol1}}. Comparison with experimental data, where available, was also provided in the table. We find that, in most cases, the lattice parameters are within 1\% of experimental (see  Supplementary material, Table S1). The pictorial representation of how experimental lattice parameters compare with computational ones is given in Fig. {\ref{fig1}}. The figure reveals good agreement between experiment and computations. We thus conclude that our  computational approach provide reliable structural predictions. The ground state structures are available from Ref.\cite{githubRelax}.

Note, that we also augmented our list of HOIPs with the following structures DMA-Zn, DMA-Co, HONH$_3$-Fe and Pna2$_1$ phase of HAZ-Mg, which so far have not been reported experimentally. Such structures were obtained by replacing Mn in DMA-Mn with Zn or Co, Mn in HONH$_3$-Mn with Fe, and Zn  in HAZ-Zn with Mg,   followed by full structural relaxation. These hypothetical structures are underscored in Table \ref{tab:exp}. 
Majority of the fully relaxed HOIPs structures retained their experimental space groups. However, there were some exceptions. 
The experimental structures of MA-Co are available in both Pnma and P2$_1$/c phases whereas experimental structures of MA-(Mn,Zn) are available only in Pnma phase. Our computations predicted that MA-(Mn,Zn,Co) are mechanically unstable in Pnma phase while P2$_1$/c phase of MA-Co is mechanically stable. To ensure mechanical stability of MA-(Mn,Zn) we deformed the Pnma structure along the eigenvector associated with negative value of $C_{44}$ and subjected such deformed structure to full structural relaxation, which resulted in  P2$_1$/c structure. It is therefore plausible, that these materials may undergo another structural phase transition to P2$_1$/c phase at low temperatures. Both P2$_1$/c and Pnma phases of MA-(Mn,Zn,Co) are reported in Table \ref{table:pol1} and Ref. \cite{githubRelax}. However, dielectric and mechanical properties were calculated from P2$_1$/c phase of the structures.

It has been reported in previous experimental studies that at low temperature, DMA-Zn  crystallizes in space group {\it Cc} with no partial occupancy at N position and possesses crystal structure similar to DMA-Mn \cite{doi:10.1021/ja801952e}. However, no structural file has been provided. Therefore, we have initiated our calculation for DMA-Zn by replacing Mn with Zn in experimentally reported DMA-Mn \cite{DMA-Mn}. In case of HAZ-Mg, previous experimental study reports a non-polar crystal structure P2$_1$2$_1$2$_1$ \cite{C3QI00034F}, but a recent DFT study \cite{C5CC06190C} shows entropy driven effects are responsible for stabilizing the structure in Pna2$_1$ space group. Therefore, we have initiated our calculation for Pna2$_1$ phase of HAZ-Mg by replacing Zn with Mg in experimentally reported Pna2$_1$ phase of HAZ-Zn \cite{C3QI00034F}. 

NH$_4$-Co experimentally is reported in P6$_3$ space group at low temperature. However, relaxed structure in the same space group was found to be mechanically unstable so further relaxation resulted in P3 space group. 

For all structures with transition metal atoms we computed energies for different magnetic orderings and selected the one with the lowest energy as the ground state. It should be noted that in agreement with previous studies\cite{Partha2, ghosh2021, ghosh2022}, we find only very small differences in energy between structures with different magnetic orderings. The magnetic orderings are given in Table \ref{table:pol1}. 

\subsection{Polarization}

\begin{figure*}
    \centering
    \includegraphics[width=1\textwidth]{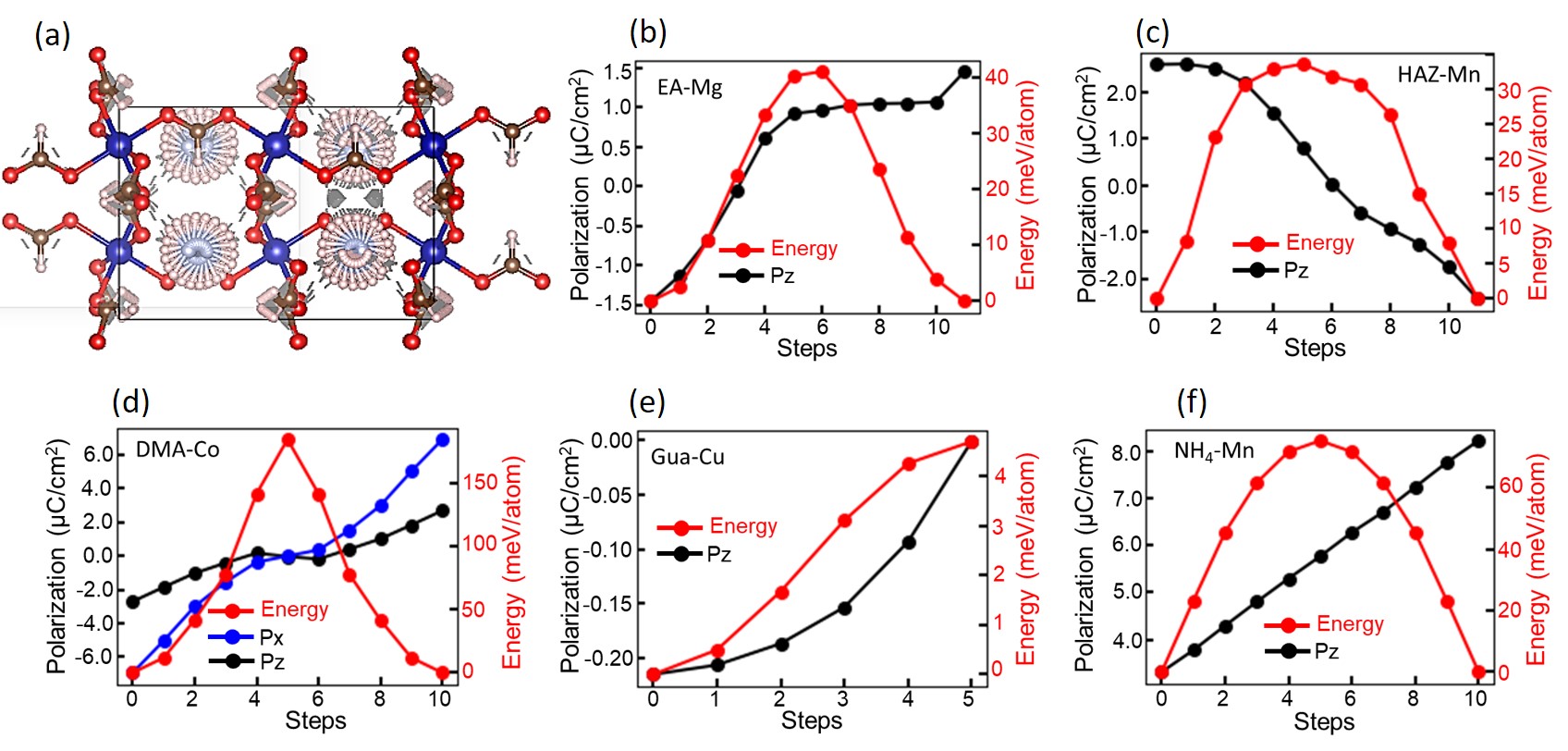}
    \caption{Structural evolution along roto-distotion path schematically shown by overlapping structures along the path (a).  Variation of polarization and energy along the path for a representative of each family, as given in the legend (b)-(f)}
    \label{fig2}
\end{figure*}

An inherent periodicity of crystal lattice makes polarization, $\mathbf{P}$, a multivalued quantity. To overcome this challenge the polarization is typically computed along a distortion path that connects polar structure to nonpolar one\cite{SPALDIN}. However, for the case of HOIP the nonpolar high symmetry structure is typically associated with partial occupancy and therefore cannot be used as a reference point. One approach to construct nonpolar phase was suggested in Ref.\cite{ Stroppa_P_reversal}. Another approach is to model experiments, where polarization is obtained from the measurement during its reversal. Such an approach was used in Ref.\cite{doi:10.1021/acs.jpcc.1c01138, Partha2}, where the polarization reversal was achieved from creation of inverted structure and generating a roto-distortion path between the structure and its inversion. The inversion was applied with respect to the inversion center of high symmetry experimental structure, where available, or with respect to B-site. The roto-distortion path consists of distortion of the framework and rotation of the A site molecule. We used same approach for EA-M, HAZ-M, DMA-M as these compounds have inversion center in their high temperature phase. 
Example of polarization evolution along such a path is given in Fig. \ref{fig2} (b) and Fig. \ref{fig2} (c).

For Gua-Cu, rotations of the Gua molecules resulted in metallic structures, which did not allow for polarization calculations. So we created a non-polar structure using pseudosymmetry module of Bilbao Crystallographic server\cite{Pseudosymmetry} and generated a distortion path between the polar and nonpolar structures. The polarization along such a path is given in Fig. \ref{fig2}(e). For NH$_4$-M family, the high temperature high symmetry structure is P6$_3$22 and does not have an inversion center. In this case, we used U2 axis of P6$_3$22 to generate the structure with reversed polarization direction. Technically, we applied the following transformation $x \rightarrow y$, $y \rightarrow x$ and $z \rightarrow -z$ on the Wykoff positions of NH$_4$-M in P6$_3$ phase. Note that for NH$_4$-Co, we report polarization for P6$_3$ phase, although it was found to be mechanically unstable in calculations. An example of polarization along such path is given in Fig. \ref{fig2} (f).

\begin{figure*}
    \centering
    \includegraphics[width=0.8\textwidth]{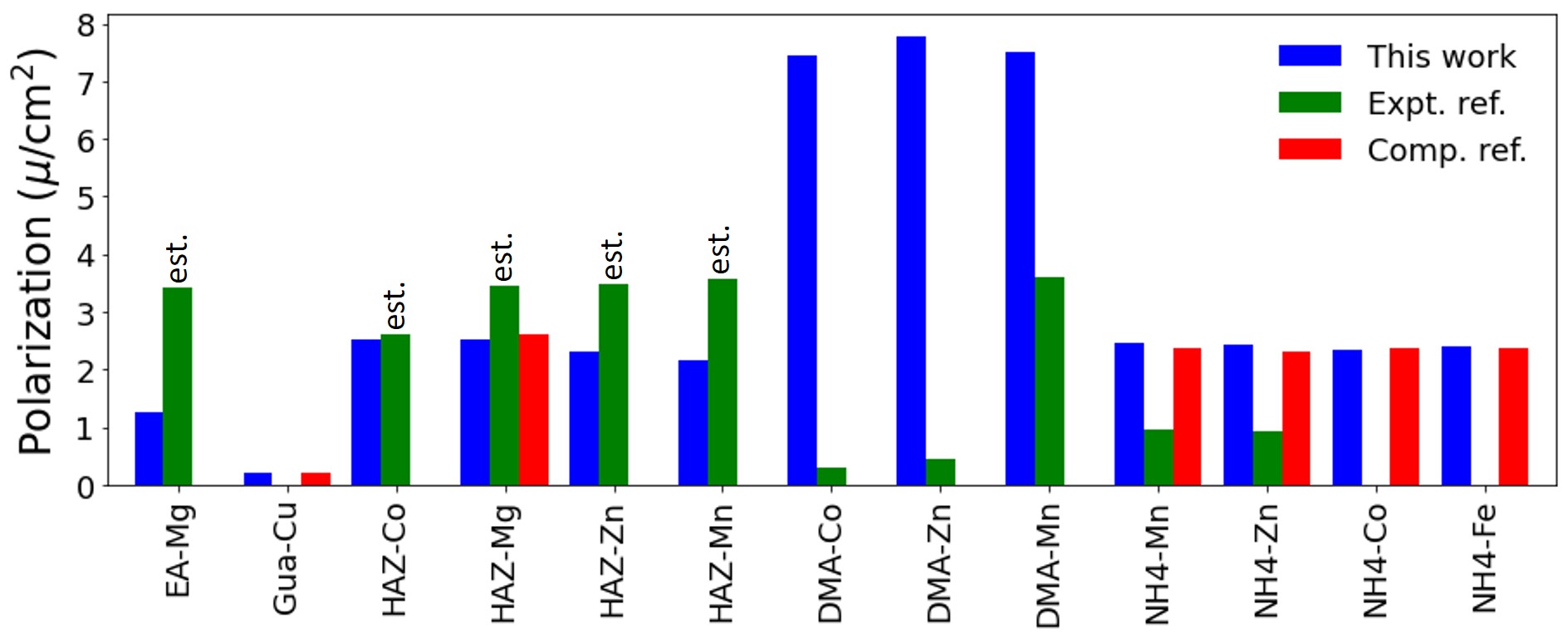}
    \caption{Comparison of our computational polarization values with experimental and computational results from  the literature\cite{C3QI00034F, C8RA00799C, doi:10.1021/acs.jpclett.0c02943, C6CP03414D, doi:10.1021/ja206891q, 10.1002/chem.201303425, doi:10.1021/ic4020702, C6CE01891B, 10.1002/chem.201303425}. Note, "est." indicates that the polarization was estimated from the separation between positive and negative charge.}
        \label{fig3}
\end{figure*}

Polarizations along the roto-distortion paths for all polar materials studied are given in Fig. S1 of Supplementary material, while the associated structures are given in Ref.\cite{githubRelax}. The Figures also report the energy along the path. The energies  are not likely to be physical as no optimization has been performed. However, they do reveal two minima, that is double-well potential. The typical barrier height is below 200~meV/atom which is considered surmountable\cite{Zunger_ferro}.
The comparison of our results with experimental and computationally predicted values available from the literature can be found in Table \ref{tab:exp-pol1} and Fig. \ref{fig3}. We find excellent agreement between our computational data and computational data from the literature. However, there exist discrepancies with experimental data. This could be attributed to the difference in temperature, and in some cases in phase, the difference in the direction of measurement. In our case we report the value along the polar direction. 
The data reveal that the polarization values for the formate family is in the range of 0.2-7.8 $\mu$C/cm$^2$ with largest values found in DMA-M. The values are a factor of ten lower than the ones for prototypical oxide ferroelectrics including BaTiO$_3$ and PbTiO$_3$ \cite{Database}.

\subsection {Piezoelectric response}

\begin{figure*}
    \centering
    \includegraphics[width=0.8\textwidth]{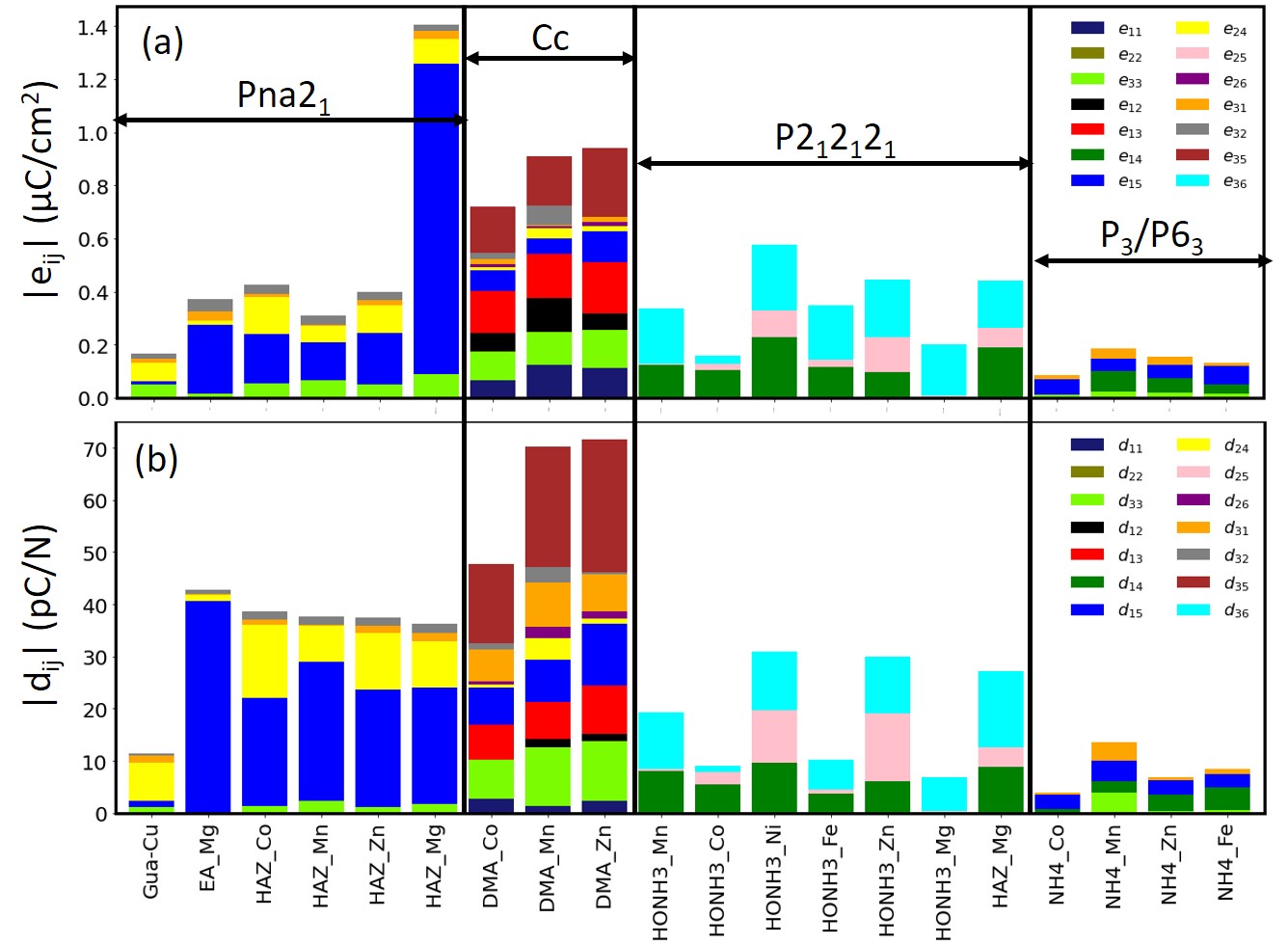}
    \caption{Comparative view of the components of the (a) piezoelectric stress and (b) piezoelectric strain tensors}
    \label{fig4}
\end{figure*}

The independent components of piezoelectric tensors, e$_{ij}$ and d$_{ij}$, which are allowed by symmetry are given in Table \ref{Table:Stress1} and Table \ref{Table:Strain1}, respectively. Figure \ref{fig4} provides comparative picture. For the formates with  Pna2$_1$ space group, we mostly find largest values for e$_{15}$ and d$_{15}$ components of the tensor. For materials in Cc and P2$_{1}$2$_{1}$2$_{1}$ space groups, the largest components are e$_{35}$ (d$_{35}$) and e$_{36}$ (d$_{36}$) respectively, and can reach 0.26~C/m$^2$ (25.36~pC/N) and 0.18~C/m$^2$ (14.64~pC/N) in DMA-Zn and HAZ-Mg, respectively. The longitudinal coefficients along the crystallographic directions, e$_{ii}$ and d$_{ii}$, $i=$1, 2, 3, range from 0.01 to 0.14~C/m$^2$ and 0.01 to 11.46~pC/N, respectively, with the largest of these values belonging to DMA-Zn. The transverse coefficients e$_{ij}$ and d$_{ij}$, $i,j=$1, 2, 3 are in the range 0.00 to 0.20~C/m$^2$ and 0.07 to 9.15~pC/N, respectively, with the largest values belonging to DMA-Zn.

\begin{figure*}
    \centering
    \includegraphics[width=0.8\textwidth]{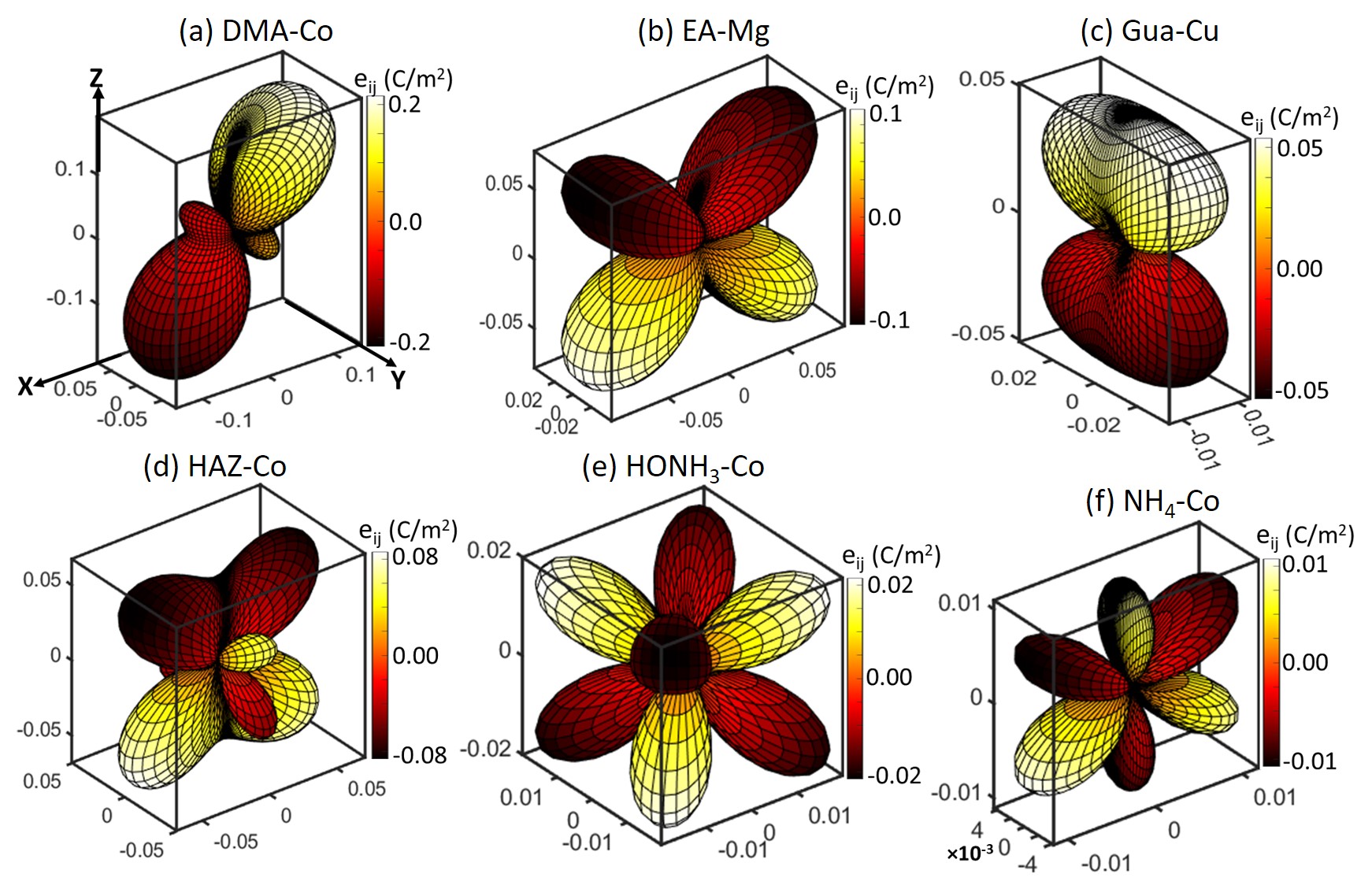}
    \caption{Piezoelectric stress surface for a representative from each family, as indicated in the titles.}
    \label{fig5}
\end{figure*}

\begin{figure*}
    \centering
    \includegraphics[width=0.8\textwidth]{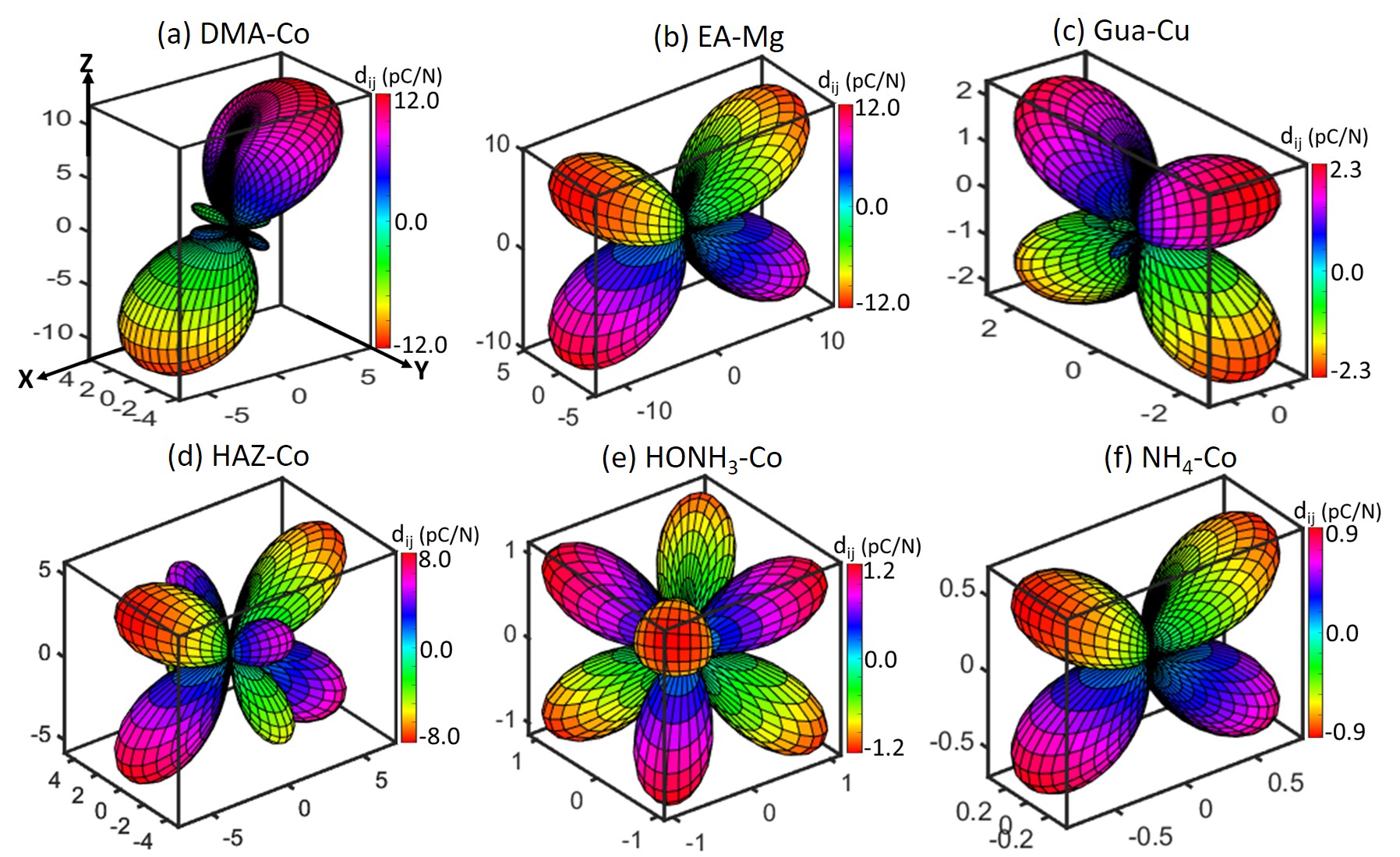}
    \caption{Piezoelectric strain surface for a representative from each family, as indicated in the titles.}
    \label{fig6}
\end{figure*}

The directional dependence of the longitudinal piezoelectric stress and strain responses was analyzed using MTex\cite{MTEX} and is presented in Fig. \ref{fig5} and Fig. \ref{fig6}, respectively, for a representative material in each family. For all materials we find response to be highly anisotropic. The longitudinal piezoelectric stress coefficient can reach 0.22 C/m$^2$ in DMA-Co in $\langle\frac{1}{2}, 0, \frac{\sqrt{3}}{2}\rangle$ direction, while the  strain coefficient can reach 12.93~pC/N in the vicinity of $\langle 1/2, 0, 1\rangle$ direction. 
3D visualizations of the piezoelectric stress/strain surfaces for the rest of the materials are given in Fig. S2 and Fig. S3 in the Supplementary material. 

Thus, our data indicate that the intrinsic piezoelectric strain response in the formate family can reach 26.7~pC/N (in HAZ-Mn) for the shear stress component  and 21.12~pC/N (in DMA-Zn along $\langle\frac{1}{2}, 0, \frac{\sqrt{3}}{2}\rangle$ direction) for the longitudinal one. DMA family exhibits the best values.

\subsection{Dielectric response}
\begin{figure*}
    \centering
    \includegraphics[width=0.8\textwidth]{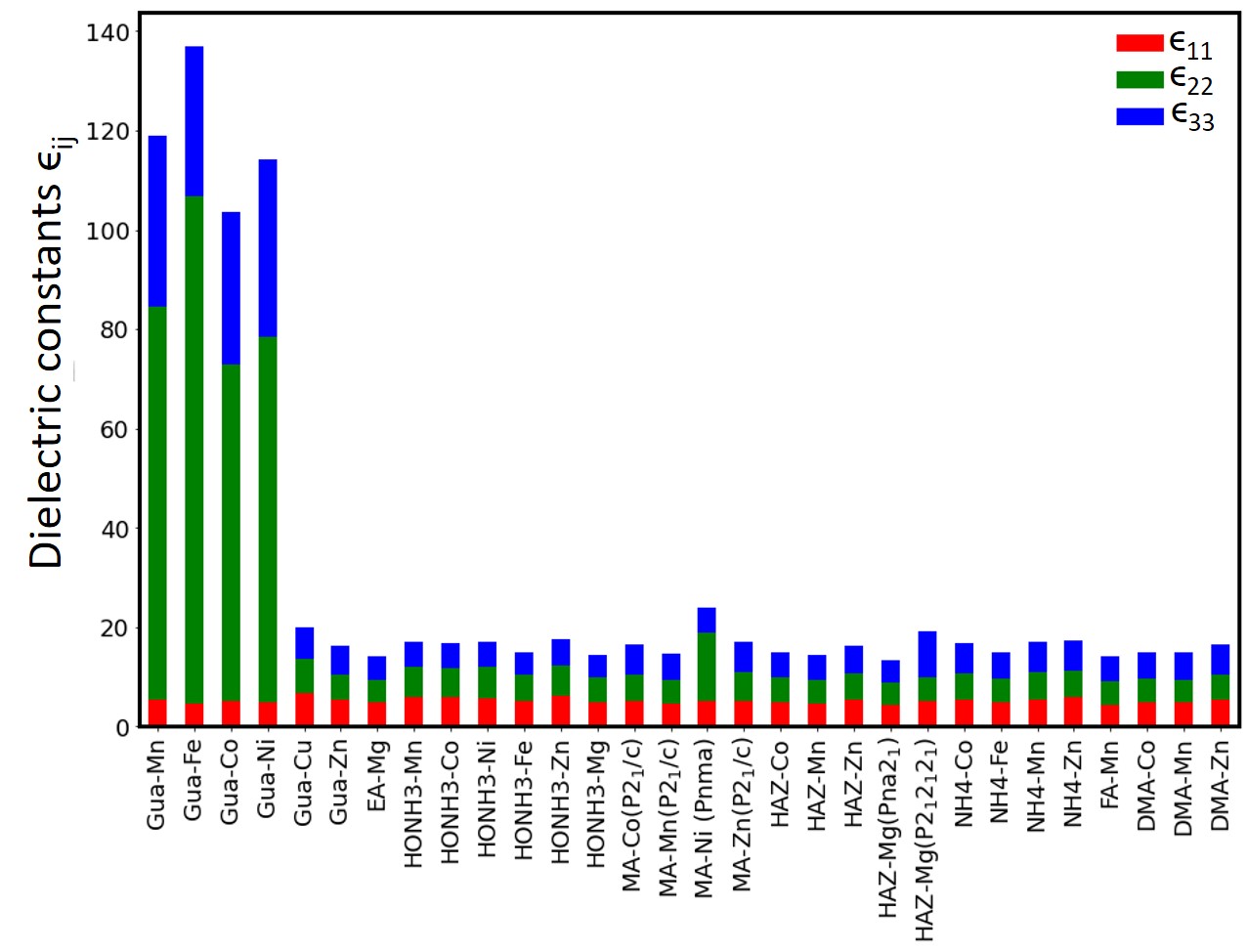}
    \caption{Comparative view of the components of the dielectric tensor.}
    \label{fig7}
\end{figure*}

The symmetry allowed components of the dielectric tensor are reported in Table \ref{Table:Dielectric1}. The typical value is 5.  
However, computations predict Gua-M to exhibit distinctively high values,  up to  100.00, comparable in order of magnitude to dielectric constants  of BaTiO$_3$\cite{BTO_D1}. The comparative view  of the dielectric constants is given in Fig. \ref{fig7}, which confirm that Gua-M family exhibits largest response. The nature of such unusual response deserves further investigation.  

\subsection{Mechanical Properties}

\begin{figure*}
    \centering
    \includegraphics[width=0.8\textwidth]{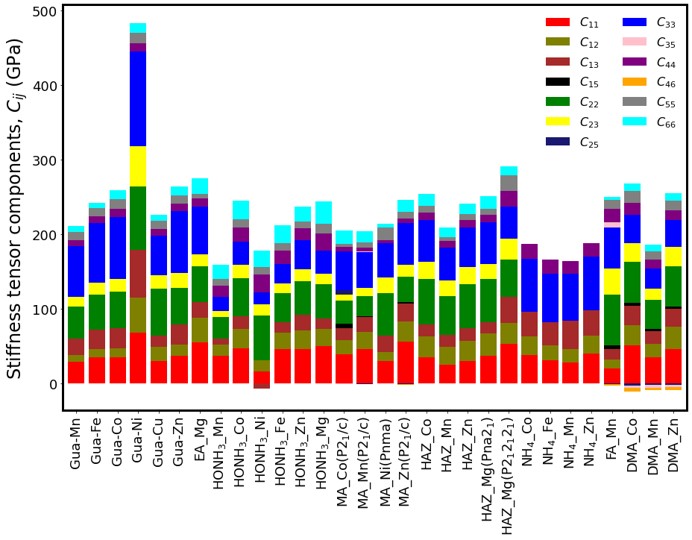}
    \caption{Comparative view of the components of the stiffness tensor (C$_{ij}$).}
    \label{fig8}
\end{figure*}

Mechanical properties describe the materials response  to external mechanical stimuli, such as   pressure, stress or strain. 
The independent components of stiffness tensors are  given  in Table \ref{Table:Elastic1}. They satisfy the Bohn conditions for the elastic stability \cite{born_1940,Bohn_edition} as checked by VASPKIT\cite{VASPKIT}. The typical diagonal elements are in the range 3.3 to 127.0. Comparative view of the stiffness tensor components among all the formates is given in Fig. \ref{fig8}. 
We computed  average Bulk modulus ($B$), Young modulus ($E$), shear modulus ($G$), Poisson's ratio ($\nu$) and  Cauchy's pressure (CP) for bulk polycrystals within the Hills' approximation as implemented in VASPKIT\cite{VASPKIT, Voigt_1928, Hill_1952, Reus_1929, VASPkit1, VASPkit2, VASPkit3} and reported them in Table \ref{Table:Mech1}. The values compare well with the experimental results, listed in Table \ref{tab:Exp_elastics}.

\begin{figure*}
    \centering
    \includegraphics[width=0.75\textwidth]{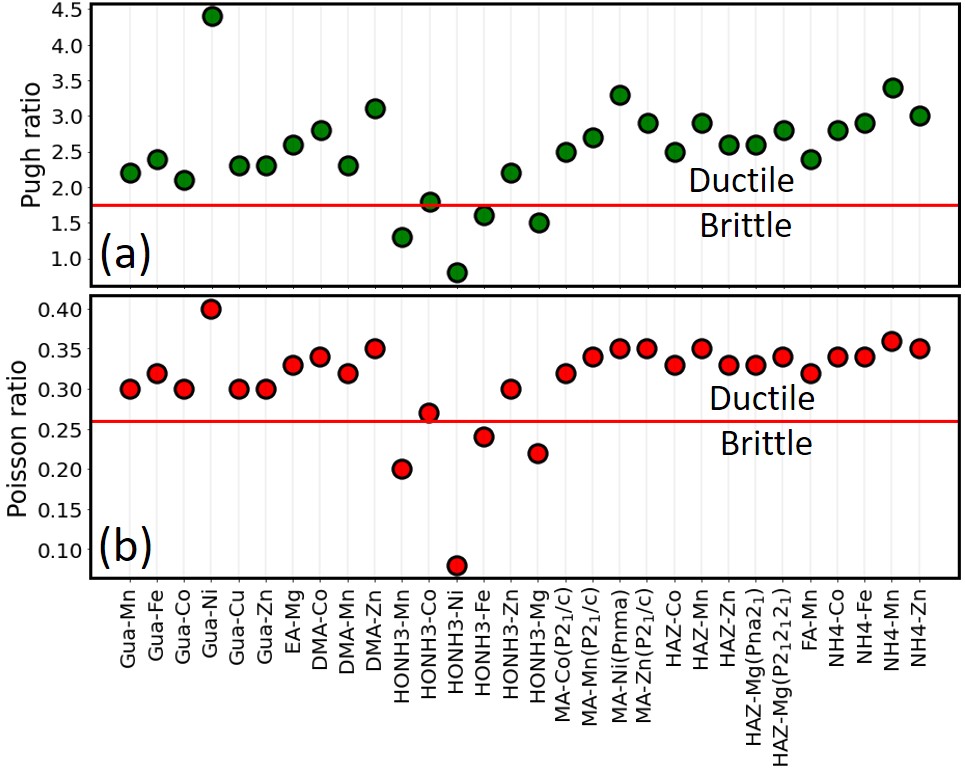}
    \caption{ Pugh (a) and Poisson (b) ratios of formates studied in this work.}
    \label{fig9}
\end{figure*}

Poisson's ratio, defined as the ratio of transverse compressive strain to longitudinal tensile strain, and Pugh's ratio, commonly expressed as $B/G$ ratio, can be used to characterize ductility or brittleness of crystals. The former one typically ranges from 0.0 to 0.5. Ductility-brittleness border line is usually drawn at Poisson ratio of 0.26 and at Pugh ratio of 1.75 \cite{Poisson1, Poisson4}. As shown in Fig. \ref{fig9}, most of the formates studied in this work are ductile and therefore are able to withstand large stresses and exhibit malleability. 

\begin{figure*}
    \centering
    \includegraphics[width=0.8\textwidth]{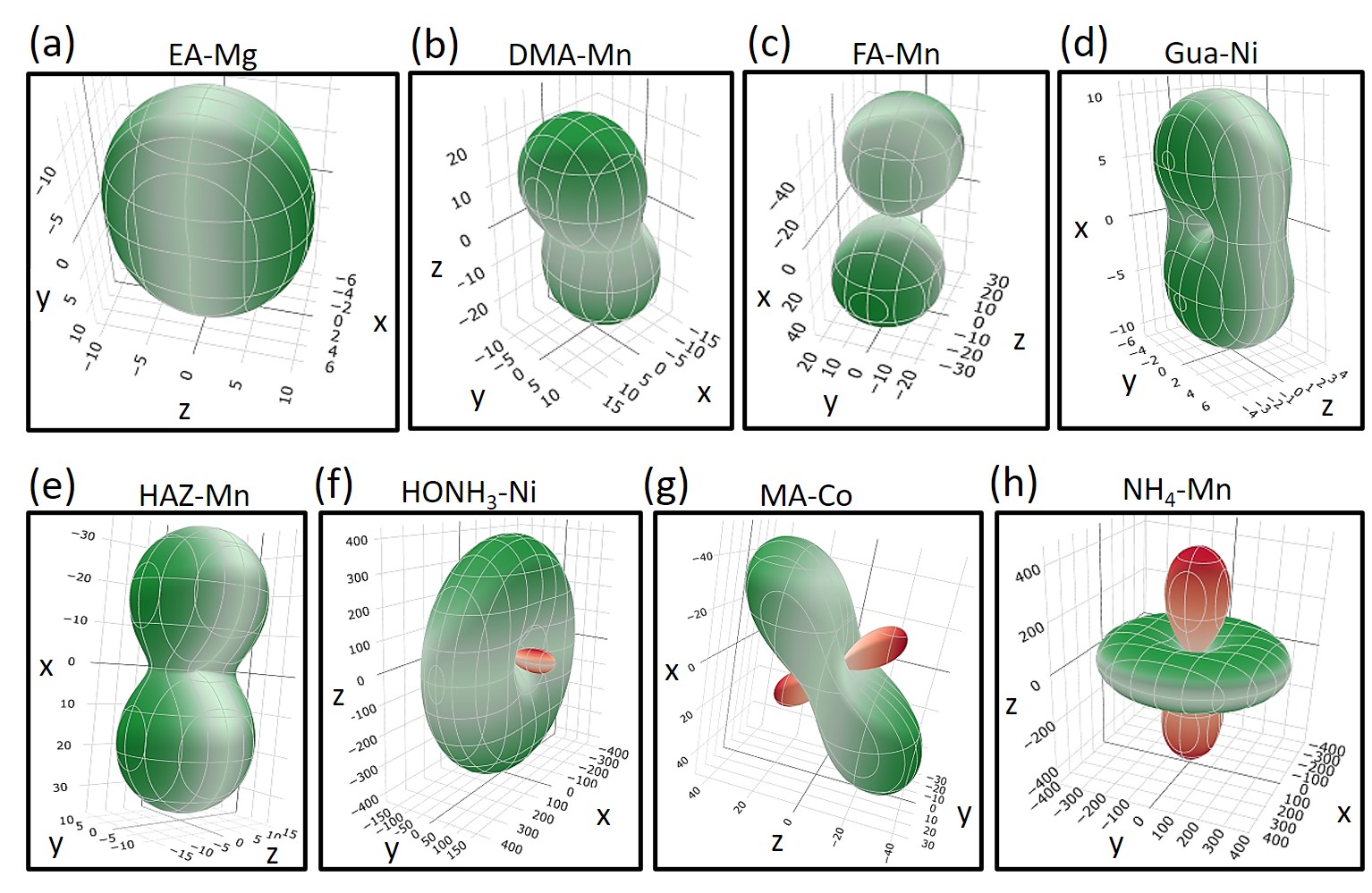}
    \caption{3D plots of linear compressibility for a representative material in each family as given in the titles. Green and red colors correspond to positive and negative values, respectively.}
    \label{fig10}
\end{figure*}

Figure \ref{fig10} shows the directional dependence of linear compressibility, defined as linear expansion or compression of materials upon application of isotropic pressure. Interestingly, the data predict that a few formates have negative values (indicated by red color) and some exhibit nearly zero values. For example, HONH$_3$-Ni, MA-Co and NH$_4$-Mn exhibit negative values along $\langle 0, 1, 0\rangle$, $\langle $-$0.6447, 0, $-$0.7644\rangle$ and $\langle0, 0, 1\rangle$ directions, respectively. Directional dependence of linear compressibility for other materials are presented in supplementary material Fig. S4. Previously negative linear compressibility was predicted for HAZ-Co, HAZ-Mn, HAZ-Fe and NH$_4$-Zn \cite{ghosh2021, ghosh2022, doi:10.1021/ja305196u} and explained on the basis of strut-hinge model\cite{strut1, Strut2}.

\begin{table*}
    \centering
    \caption{Piezoelectric stress constants $e_{ij}$ in C/m$^2$. Materials which do not have experimentally reported structure are underscored.}
    \begin{ruledtabular}
    \begin{tabular}{ccccccccccc}
    &e$_{15}$&e$_{24}$&e$_{31}$&e$_{32}$& e$_{33}$\\
    Gua-Cu & $-$0.011& 0.069& $-$0.017& 0.018& 0.051\\
    \hline
    & e$_{15}$& e$_{24}$& e$_{31}$ & e$_{32}$ & e$_{33}$\\
    EA-Mg &$-$0.261&$-$0.014& $-$0.035& $-$0.045&$-$0.015\\
    \hline
    & e$_{11}$& e$_{12}$& e$_{13}$ & e$_{15}$ & e$_{24}$& e$_{26}$& e$_{31}$& e$_{32}$& e$_{33}$&e$_{35}$\\
    \underline{DMA-Co} &0.065& 0.071& 0.160& 0.077& 0.012& $-$0.009&$-$0.020&0.025&0.107&0.175\\
    DMA-Mn &0.124&0.129&0.165&0.058&0.041&0.006&$-$0.004&0.073&0.122&0.187 \\
    \underline{DMA-Zn} &0.112&0.062&0.196&0.114&0.019&$-$0.017&0.020&0.001&0.141&0.258\\
    \hline
    &e$_{14}$& e$_{25}$& e$_{36}$ &\\
    HONH$_3$-Mn &$-$0.124 & 0.003 &$-$0.211\\
    HONH$_3$-Co &$-$0.102&$-$0.027&$-$0.031\\
    HONH$_3$-Ni &$-$0.226&0.103&0.247\\
    \underline{HONH$_3$-Fe} &$-$0.065&$-$0.004&$-$0.135\\
    HONH$_3$-Zn&$-$0.097&$-$0.130&$-$0.218\\
    HONH$_3$-Mg&$-$0.005&$-$0.002&$-$0.193\\
    \hline
    &e$_{15}$& e$_{24}$& e$_{31}$ & e$_{32}$ & e$_{33}$\\
    HAZ-Co&$-$0.185& 0.138& $-$0.012&0.035&$-$0.054\\
    HAZ-Mn&$-$0.143&0.060&0.004&0.037&$-$0.067\\
    HAZ-Zn&$-$0.194&0.104&$-$0.021&0.028&$-$0.050\\
    \underline{HAZ-Mg (Pna2$_1$)}&$-$1.172&0.090&$-$0.032&0.022&$-$0.088 \\
    &e$_{14}$& e$_{25}$& e$_{36}$\\
    HAZ-Mg (P2$_1$2$_1$2$_1$)&$-$0.190&$-$0.074&$-$0.177 \\
    \hline
    & e$_{14}$& e$_{15}$&e$_{31}$&e$_{33}$\\
    NH$_4$-Co&0.001&$-$0.057&0.017&0.011\\
    NH$_4$-Fe&0.078&$-$0.046&0.037&0.023\\
    NH$_4$-Zn&0.055&$-$0.049&0.031&0.019\\
    NH$_4$-Mn&0.034&$-$0.069&$-$0.013&$-$0.015\\
    \end{tabular}
    \end{ruledtabular}
    \label{Table:Stress1}
\end{table*}

\begin{table*}
    \centering
    \caption{Piezoelectric stress constants $d_{ij}$ in pC/N. Materials which do not have experimentally reported structure are underscored.}
    \begin{ruledtabular}
    \begin{tabular}{ccccccccccccccc}
    &d$_{15}$&d$_{24}$&d$_{31}$&d$_{32}$&d$_{33}$\\
    Gua-Cu&$-$1.05&7.36&$-$1.41&0.39&1.23 \\
    \hline
    &d$_{15}$&d$_{24}$&d$_{31}$ & d$_{32}$ &d$_{33}$\\
    EA-Mg &$-$40.55&$-$1.24&$-$0.14&$-$0.84&0.01\\
    \hline
    &d$_{11}$&d$_{12}$&d$_{13}$ & d$_{15}$ &d$_{24}$&d$_{26}$&d$_{31}$&d$_{32}$&d$_{33}$&d$_{35}$\\
    \underline{DMA-Co} &$-$2.77&0.18&6.52&7.21&0.59&$-$0.57&$-$6.19&1.19&7.37&15.05\\
    DMA-Mn &$-$1.29&1.58&7.11&8.16&4.04&2.19&$-$8.56&2.81&11.29&23.16\\
    \underline{DMA-Zn}&$-$2.34&$-$1.39&9.15&11.85&1.05&$-$1.32&$-$7.10&$-$0.50&11.46&25.36\\
    \hline
    &d$_{14}$& d$_{25}$& d$_{36}$ &\\
    HONH$_3$-Mn &$-$8.06&0.33&$-$11.01\\
    HONH$_3$-Co &$-$5.47&$-$2.35&$-$1.25\\
    HONH$_3$-Ni&$-$9.65&9.97&11.38\\
    \underline{HONH$_3$-Fe} &$-$3.66&$-$0.75&$-$5.85\\
    HONH$_3$-Zn &$-$6.03&$-$13.16&$-$10.86\\
    HONH$_3$-Mg &$-$0.22&$-$0.16&$-$6.47\\

    \hline
    &d$_{15}$& d$_{24}$& d$_{31}$& d$_{32}$& d$_{33}$\\
    HAZ-Co &$-$20.73&14.08&$-$1.01&1.54&$-$1.32\\
    HAZ-Mn&$-$26.72&6.90&$-$0.07&1.71&$-$2.33\\
    HAZ-Zn&$-$22.48&10.71&$-$1.41&1.62&$-$1.24\\
    \underline{HAZ-Mg (Pna2$_1$)}&$-$22.40&8.86&$-$1.53&1.73&$-$1.74\\
    &d$_{14}$& d$_{25}$& d$_{36}$\\
    HAZ-Mg (P2$_1$2$_1$2$_1$)&$-$8.91&$-$3.62&$-$14.64\\
    \hline
    &d$_{14}$&d$_{15}$& d$_{31}$ & d$_{33}$\\
    NH$_4$-Co&0.49&$-$2.90&0.37&$-$0.20\\
    NH$_4$-Mn&1.99&$-$4.09&$-$3.55&4.00\\
    NH$_4$-Zn&3.07&$-$2.73&0.71&$-$0.42 \\
    NH$_4$-Fe&4.30&$-$2.47&1.05&$-$0.63\\
    \end{tabular}
    \end{ruledtabular}
    \label{Table:Strain1}
\end{table*}       

\begin{table*}
    \centering
    \caption{Components of stiffness tensor $C_{ij}$ in GPa. Materials which do not have experimentally reported structure are underscored.}
    \begin{ruledtabular}
    \begin{tabular}{cccccccccccccc}
    & C$_{11}$& C$_{12}$& C$_{13}$ & C$_{22}$& C$_{23}$& C$_{33}$& C$_{44}$& C$_{55}$& C$_{66}$\\

    Gua-Mn  & 28.6 & 9.5 & 22.3 & 42.6 & 13.2 & 68.2 & 7.94 & 10.6 & 8.3\\
    Gua-Fe &34.6&11.2&26.5&46.9&15.3 &80.0&9.3&11.6&6.4 \\
    Gua-Co &34.5& 12.3 & 26.7 & 49.9 & 17.0 & 82.9 & 10.3 & 13.0 & 11.9\\
    Gua-Ni  &68.4 & 46.7 & 64.2 & 84.6& 53.6 & 127.0 & 11.0 & 14.4 & 13.0\\
    Gua-Cu &29.7& 19.7& 14.3& 63.6& 17.4& 52.8& 9.4&  10.6& 8.9 \\
    Gua-Zn&36.6&15.3&27.0&49.4&19.5&83.1&10.3&10.9&12.0\\
    \hline
    & C$_{11}$& C$_{12}$& C$_{13}$ & C$_{22}$& C$_{23}$& C$_{33}$& C$_{44}$& C$_{55}$& C$_{66}$\\
    EA-Mg &55.3&32.3& 20.9& 48.5& 15.9& 63.6& 11.2& 6.4& 21.2\\
    \hline
    & C$_{11}$& C$_{12}$& C$_{13}$ & C$_{15}$& C$_{22}$& C$_{23}$& C$_{25}$& C$_{33}$& C$_{35}$& C$_{44}$& C$_{46}$& C$_{55}$& C$_{66}$\\
    \underline{DMA-Co}&50.6& 27.7& 25.2& 4.9& 54.8& 24.8& $-$3.4& 37.8 &$-$3.0 &16.3& $-$4.8&15.4&10.1\\
    DMA-Mn&34.9& 18.3 &17.1& 2.2& 39.6 &15.2& $-$2.3&26.7&$-$3.3 &12.0& $-$3.3&10.8&8.8\\
    \underline{DMA-Zn}&46.4&29.2&24.2&3.4&54.2&25.4&$-$2.2&35.7&$-$3.3&13.9&$-$3.3&12.6&9.9\\
    \hline
    & C$_{11}$& C$_{12}$& C$_{13}$ & C$_{22}$& C$_{23}$& C$_{33}$& C$_{44}$& C$_{55}$& C$_{66}$\\
    HONH$_3$-Mn &36.9& 15.2& 7.8 &29.3& 8.0& 18.6& 15.4& 9.0& 19.2\\
    HONH$_3$-Co &46.9 & 25.6 & 17.9 & 50.9 & 18.0 & 31.1 & 18.6 & 11.3 & 25.0 \\
    HONH$_3$-Ni & 16.1 & 15.0 & $-$6.6 & 59.8 & 15.3 & 16.1 & 23.4 & 10.4 & 21.7 \\
    \underline{HONH$_3$-Fe} & 46.0 & 21.8 & 13.7 & 39.1 & 12.9 & 26.2 & 17.9 & 10.8 & 23.1 \\
    HONH$_3$-Zn &45.6&25.1&21.4&44.8&16.5&38.4&16.0&9.0&20.1\\
    HONH$_3$-Mg &50.2&22.5&14.2&45.6&14.4&31.2&23.2&12.6&29.8\\
    \hline
    & C$_{11}$& C$_{12}$& C$_{13}$ & C$_{22}$& C$_{23}$& C$_{33}$& C$_{44}$& C$_{55}$& C$_{66}$\\
    MA-Ni (Pnma) & 29.7 & 12.2 & 22.2 & 56.7& 21.2& 46.3 & 3.6& 17.3 & 4.9 \\
    &C$_{11}$& C$_{12}$& C$_{13}$ & C$_{15}$ &C$_{22}$& C$_{23}$&C$_{25}$& C$_{33}$&C$_{35}$& C$_{44}$&C$_{46}$& C$_{55}$& C$_{66}$\\
    MA-Co (P2$_1$/c)& 39.1& 18.6 &16.3&6.0&30.8&8.6&5.2&52.1&$-$0.8&6.2&0.5& 3.3&18.5\\
    \underline{MA-Zn (P2$_1$/c)}  &55.9&27.1&24.3&1.8&33.8&16.4&$-$1.4&55.2&0.6&6.2&$-$0.7&8.2&16.1\\
    \underline{MA-Mn (P2$_1$/c)}  &46.4&22.7&19.7&1.3&26.4&11.6&$-$1.2&47.8&0.8&4.9&$-$0.3&7.0&15.5\\  

    \hline
    &C$_{11}$& C$_{12}$& C$_{13}$ & C$_{22}$& C$_{23}$& C$_{33}$& C$_{44}$& C$_{55}$& C$_{66}$\\
   
    HAZ-Co&35.0&28.4&15.2& 61.2& 23.3& 56.0& 9.8& 9.0& 15.8\\
    HAZ-Mn&25.0&24.4&15.5&52.0&21.5&43.9&8.7& 5.3&13.0\\
    HAZ-Zn&30.4&26.6&16.7&58.8&23.9&52.4&9.7& 8.7&13.7\\ 
    \underline{HAZ-Mg (Pna2$_1$)}&36.8& 29.7& 15.3& 58.3& 19.5& 56.6& 10.1& 7.7& 16.9\\
    HAZ-Mg (P2$_1$2$_1$2$_1$)&53.3&28.0&35.1&49.1& 28.5&42.7&21.3&20.5&12.1\\
    \hline
    &C$_{11}$& C$_{12}$& C$_{13}$ & C$_{15}$ &C$_{22}$& C$_{23}$&C$_{25}$& C$_{33}$&C$_{35}$& C$_{44}$&C$_{46}$& C$_{55}$& C$_{66}$\\    
    FA-Mn&19.6&11.9&14.1&5.7&67.2&35.8&$-$0.9&54.4&7.1&18.0&$-$2.1&12.5&3.4\\
    \hline
    &C$_{11}$& C$_{12}$& C$_{13}$& C$_{33}$& C$_{44}$\\
    NH$_4$-Co&38.4&24.1&33.2&71.5&20.2\\
    NH$_4$-Fe&30.8& 20.6& 30.4& 65.6& 18.1\\
    NH$_4$-Mn& 28.3 & 17.7 & 37.7 & 63.1 & 17.0\\
    NH$_4$-Zn&39.9&24.2&34.3&71.4&18.0\\
    \end{tabular}
    \end{ruledtabular}
    \label{Table:Elastic1}
\end{table*}

\begin{table*}
    \centering
    \caption{Dielectric constants. Materials which do not have experimentally reported structure are underscored.}
    \begin{ruledtabular}
    \begin{tabular}{ccccccc}
    & $\epsilon_{11}$& $\epsilon_{22}$&$\epsilon_{33}$\\
    \hline
    Gua-Mn  &5.31&79.19&34.42 \\
    Gua-Fe &4.62&102.12&30.26\\
    Gua-Co  &5.26&67.72&30.55\\
    Gua-Ni  &4.94&73.59&35.51\\
    Gua-Cu &6.79&6.85&6.26\\
    Gua-Zn  &5.33&5.20&5.68\\
    \hline
    & $\epsilon_{11}$& $\epsilon_{22}$&$\epsilon_{33}$\\
    EA-Mg &4.82 &4.63&4.67\\
    \hline
    & $\epsilon_{11}$&$\epsilon_{22}$&$\epsilon_{33}$&$\epsilon_{13}$&Expt&Ref.\\
    \underline{DMA-Co} &4.92&4.61&5.35&0.33&\\
    DMA-Mn &4.93&4.53&5.53&0.41&3 -- 6&\cite{DMA-Mn-exp}\\
    \underline{DMA-Zn} &5.50&4.98&6.00&0.41&8 -- 10&\cite{DMA-Zn-die}\\
    \hline
    & $\epsilon_{11}$& $\epsilon_{22}$&$\epsilon_{33}$\\
    HONH$_3$-Mn&5.90&6.04&5.21\\
    HONH$_3$-Co&5.84&5.93&4.95\\
    HONH$_3$-Ni&5.56&6.48&5.08\\
    \underline{HONH$_3$-Fe}&5.14&5.29&4.46\\
    HONH$_3$-Zn&6.26&6.04&5.15\\
    HONH$_3$-Mg&4.84&5.01&4.43\\
    \hline
    & $\epsilon_{11}$& $\epsilon_{22}$&$\epsilon_{33}$&$\epsilon_{13}$\\
    MA-Co (P2$_1$/c) &5.21&5.34&5.89&0.22\\
    \underline{MA-Zn (P2$_1$/c)} &5.17&5.83&6.06&$-$0.20\\
    \underline{MA-Mn (P2$_1$/c)}&4.62&4.85&5.29&$-$0.13\\
    MA-Ni (Pnma)&5.21&13.69&5.11\\

    \hline
    & $\epsilon_{11}$& $\epsilon_{22}$&$\epsilon_{33}$\\
    \hline
    HAZ-Co&4.87&5.05&5.03\\
    HAZ-Mn&4.66&4.84&4.83\\
    HAZ-Zn&5.30&5.41&5.49\\ 
    \underline{HAZ-Mg (Pna2$_1$)}&4.31&4.65&4.50\\
    HAZ-Mg (P2$_1$2$_1$2$_1$)&5.17&4.75&9.29\\
    \hline
    & $\epsilon_{11}$&$\epsilon_{22}$&$\epsilon_{33}$&$\epsilon_{13}$\\
    FA-Mn&4.36&4.72&5.07&0.27\\
    \hline
    & $\epsilon_{11}$& $\epsilon_{22}$&$\epsilon_{33}$\\
    NH$_4$-Co&5.38&5.38&6.04\\
    NH$_4$-Fe&4.77&4.77&5.28\\
    NH$_4$-Mn&5.51&5.51&5.94\\
    NH$_4$-Zn&5.28&5.28&6.17\\
    \end{tabular}
    \end{ruledtabular}
    \label{Table:Dielectric1}
\end{table*}

\begin{table*}
    \centering
    \caption{Mechanical properties of formates. $B$, $E$ and $G$ are average bulk, Young and shear moduli of bulk polycrystal, respectively. B/G, $\nu$, CP and LC$_{min}$ are Pugh's ratio, Poisson's ratio, Cauchy's pressure and minimum linear compressibility, respectively. Materials which do not have experimentally reported structure are underscored}
    \begin{ruledtabular}
    \begin{tabular}{cccccccccc}
    & B (GPa) & E (GPa) & G (GPa) & B/G & $\nu$ & CP (GPa) & LC$_{min}$ (TPa$^{-1}$)\\

    Gua-Mn &23.5 &28.2 &10.8& 2.2 &0.30 &1.5 &2.41\\
    Gua-Fe &27.4 & 30.2&11.5&2.4 &0.32 &1.9 &2.18\\
    Gua-Co &29.0 &35.7 &13.8&2.1 &0.30 &2.0 &2.45\\
    Gua-Ni &64.1&40.7&14.6&4.4&0.4&35.6&0.05\\
    Gua-Cu &26.1&30.0 &11.4&2.3&0.3&10.4 &5.03\\
    Gua-Zn   &30.3&34.2&13.0&2.3&0.3&5.0&2.46\\
    \hline
    EA-Mg &33.7&34.2&12.9&2.6&0.33&21.1&6.72\\
    \hline
    \underline{DMA-Co} &32.5&31.3&11.7&2.8&0.34&11.4&6.85\\
    DMA-Mn &20.9&23.8&9.0&2.3&0.32&6.3&8.51\\
    DMA-Zn &31.7&28.2&10.4&3.1&0.35&15.3&5.36\\
    \hline
    HONH$_3$-Mn &15.3&27.5&11.5&1.3&0.20&$-$0.2 &11.33\\
    HONH$_3$-Co &26.8&36.9&14.5&1.8&0.27&7.0&7.18\\
    HONH$_3$-Ni &8.5 & 21.7&10.1&0.8 & 0.08&$-$8.4 &$-$188.08\\
    \underline{HONH$_3$-Fe} &22.0 &34.0 &13.6&1.6 &0.24 &3.9 &7.63\\
    HONH$_3$-Zn &28.0&33.8&13.0&2.2&0.30&9.1&7.51\\
    HONH$_3$-Mg &24.5&41.2&16.9&1.5&0.22&$-$0.7&8.72\\
    \hline
    MA-Co (P2$_1$/c)&19.9 &21.1 &8.0&2.5 &0.32 &12.4 &$-$32.24\\
    \underline{MA-Mn (P2$_1$/c)}&23.7 &23.4 &8.8 &2.7&0.34 &17.8&$-$0.15\\
    MA-Ni (Pnma)& 27.8&23.1 &8.5 &3.3&0.35 &8.6 &1.96\\
    \underline{MA-Zn (P2$_1$/c)} &29.7&27.5&10.2&2.9&0.35&20.9&4.22\\
    \hline
   
    HAZ-Co&30.1&31.5&11.9&2.5&0.33&18.6&1.77\\
    HAZ-Mn&24.8&23.1&8.6&2.9&0.35&15.6&$-$1.73\\
    HAZ-Zn&28.6&28.8&10.8&2.6&0.33&16.9&0.82\\ 
    \underline{HAZ-Mg (Pna2$_1$)}&30.2&31.3&11.8&2.6&0.33&19.6&3.28\\
    HAZ-Mg (P2$_1$2$_1$2$_1$)&36.2&34.7&12.9&2.8&0.34&6.6&5.28\\
    \hline
    FA-Mn&22.8&25.4&9.6&2.4&0.32&$-$6.1&1.07\\
    \hline
    NH$_4$-Co&33.9&32.0&11.9&2.8&0.34&3.9&$-$1.73\\
    NH$_4$-Fe&29.5&27.8&10.3&2.9&0.34&2.5&$-$4.50\\
    NH$_4$-Mn&18.5&15.1&5.5&3.4&0.36&0.7&$-$439.63\\
    NH$_4$-Zn&34.7&31.5&11.7&3.0&0.35&6.2&$-$2.04
    \end{tabular}
    \end{ruledtabular}
    \label{Table:Mech1}
\end{table*}

\section{Conclusion and Outlook}

In summary, we have used DFT computations to assess structural, electric, piezoelectric, and mechanical properties of  29 hybrid formate perovskites. We predict that the ground state phase of most MA-M (M = Co, Mn, Zn) formates is different from the low temperature phase reported experimentally, which suggests additional phase transitions at very low temperatures. The spontaneous polarizations range from 0.2 to 7.8~$\mu$C/cm$^2$ with the largest values being in DMA-M family. They are expected to be reversible by the electric field as the upper estimate for the energy barrier is 200 meV/atom. We also find polarization values often exceeding experimentally reported ones, which we attribute to the difference in the direction of measurement. Thus, our study could guide towards optimization of materials performance. Typical dielectric constants are typically 5.0. Nevertheless, Gua family exhibits outstandingly large values in range 4.6 -- 102.1, which, however, need to be further validated. Intrinsic piezoelectric strain and stress constants are in the range 0.1 -- 25.8~$\mu$C/cm$^2$ and 0.1 -- 26.7~pC/N, respectively. The responses were also found to be highly anisotropic. Components of elastic stiffness tensor range from 0.3 to 127.0~GPa. On the basis of Pugh and Poisson ratio we found most of the materials to be ductile. Computations predict that linear compressibility is highly anisotropic and many materials (e.g. HONH$_3$-Ni, NH$_4$-Mn, Gua-Ni and MA-Co) exhibit either zero or even negative values. All computational data are available from Ref. \cite{githubRelax}.

Our study reveals that additional investigations are needed to validate and explain  outstanding dielectric response of Gua-M formates, and large piezoelectric response of DMA-M formates, along with the large negative compressibility values for HONH$_3$-Ni and NH$_4$-Mn. Investigation on the origin of negative and/or nearly zero values of compressibility is also required.

\section{Acknowledgment}

The work is supported by the National Science Foundation under the grant EPMD-2029800. 

\section{Data availability}
The data that support the findings of this study are available from our github repository\cite{githubRelax} as well as from the corresponding author upon reasonable request.

\section{References}
\bibliography{paper}

\newpage

\onecolumngrid

\newpage

\begin{center}
   \textbf{\Large Supplementary Material}
\end{center}

  \renewcommand{\thefigure}{S\arabic{figure}}
\setcounter{figure}{0}

  \renewcommand{\thetable}{S\arabic{table}}
\setcounter{table}{0}

\begin{table}[h]
    \centering
    \caption{Computational details and errors associated with estimation of structural parameters. KEYS: ENCUT = Cut-Off Energy and K$_D$ = k-points Density}
    \begin{ruledtabular}
    \begin{tabular}{ccccccc}
    & Error-a (\%)& Error-b (\%) &Error-c (\%) &Error-V (\%) & ENCUT (eV)&K$_D$ (\AA$^{-1}$)\\
    \hline
    Gua-Mn  &$-$1.2&$-$0.6&0.7&$-$1.1&708&0.33 \\
    Gua-Fe &$-$0.8&$-$0.6&0.3&$-$1.1&850&0.27\\
    Gua-Co  &$-$0.6&$-$0.9&0.9&$-$0.6&708&0.39\\
    Gua-Ni  &$-$0.2&$-$0.1&0.9&$-$0.6&850&0.33\\
    Gua-Cu &$-$0.2&0.4&$-$0.7&$-$0.5&700&0.33\\
    Gua-Zn  &$-$1.0&$-$0.6&$-$0.9&0.7&708&0.27\\
    \hline
    EA-Mg &$-$0.9&0.0&$-$0.7&$-$1.6&700&0.42\\
    \hline
    DMA-Mn &0.0&0.9&-0.2&0.6&700&0.33\\
    \hline
    HONH$_3$-Mn&$-$0.1&1.0&$-$0.8&0.0&850&0.28\\
    HONH$_3$-Co&$-$0.1&0.8&$-$0.2&0.5&850&0.28\\
    HONH$_3$-Ni&$-$0.4&2.6&0.5&2.7&850&0.29\\
    HONH$_3$-Zn&$-$0.6&1.2&1.2&1.9&850&0.57\\
    HONH$_3$-Mg&$-$0.3&1.3&$-$1.0&0.0&708&0.29\\
    \hline
    MA-Co(Pnma)&$-$1.2&0.0&1.6&0.4&708&0.34\\
    MA-Co(P21/c)&0.9&0.2&0.6&1.6&708&0.34\\
    MA-Mn&-3.3&$-$0.2&3.1&$-$0.5&708&0.34\\
    MA-Ni&$-$0.4&0.9&2.0&2.5&708&0.34\\
    MA-Zn(Pnma)&$-$1.2&$-$0.2&0.9&$-$0.5&708&0.28\\
    \hline
    FA-Mn&$-$2.4&2.1&$-$2.1&$-$2.5&700&0.33\\
    \hline
    HAZ-Co&$-$0.2&$-$0.3&$-$0.8&$-$1.3&700&0.40\\
    HAZ-Mn&0.7&0.1&$-$0.3&0.5&700&0.34\\
    HAZ-Zn&$-$0.1&0.4&0.1&0.3&700&0.35\\ 
    HAZ-Mg (P2$_1$2$_1$2$_1$)&1.4&1.1&$-$1.4&1.1&700&0.40\\
    \hline
    NH$_4$-Co&$-$0.4&0$-$0.4&0.2&0.5&850&0.22\\
    NH$_4$-Fe&$-$0.3&$-$0.3&2.5&1.9&850&0.19\\
    NH$_4$-Mn&$-$0.9&$-$0.9&2.0&0.0&850&0.22\\
    NH$_4$-Zn&$-$0.2&$-$0.2&2.2&1.7&708&0.22\\

    \end{tabular}
    \end{ruledtabular}
    \label{table:S1}
\end{table}

\begin{table*}
    \centering
    \caption{Mechanical properties of formates. $B$, $E$ and $G$ are average bulk, Young and shear moduli of bulk polycrystal, respectively. B/G, $\nu$, CP and LC$_{min}$ are Pugh's ratio, Poisson's ratio, Cauchy's pressure and minimum linear compressibility, respectively. Materials which do not have experimentally reported structure are underscored}
    \begin{ruledtabular}
    \begin{tabular}{cccccccccc}
    & B(GPa) & E(GPa) & G(GPa) & B/G & v & CP(GPa) & ALC & LC$_{min}$(TPa$^{-1}$) & LC$_{max}$(TPa$^{-1}$)\\

    Gua-Mn &23.5 &28.2 &10.8& 2.2 &0.30 &1.5 &11.5&2.41&27.04\\
    Gua-Fe &27.4 & 30.2&11.5&2.4 &0.32 &1.9 &10.2&2.18&22.32\\
    Gua-Co &29.0 &35.7 &13.8&2.1 &0.30 &2.0 &8.5&2.45&20.87\\
    Gua-Ni &64.1&40.7&14.6&4.4&0.4&35.6&215.6&0.05&10.47\\
    Gua-Cu &26.1&30.0 &11.4&2.3&0.3&10.4 &5.0&5.03&25.32\\
    Gua-Zn   &30.3&34.2&13.0&2.3&0.3&5.0&8.2&2.46&20.04\\
    \hline
    EA-Mg &33.7&34.2&12.9&2.6&0.33&21.1&1.9&6.72&12.76\\
    \hline
    \underline{DMA-Co} &32.5&31.3&11.7&2.8&0.34&11.4&2.6&6.85&17.54\\
    DMA-Mn &20.9&23.8&9.0&2.3&0.32&6.3&3.2&8.51&26.90\\
    DMA-Zn &31.7&28.2&10.4&3.1&0.35&15.3&3.6&5.36&19.55\\
    \hline
    HONH$_3$-Mn &15.3&27.5&11.5&1.3&0.20&$-$0.2 &3.7&11.33&41.90\\
    HONH$_3$-Co &26.8&36.9&14.5&1.8&0.27&7.0&3.2&7.18&23.08\\
    HONH$_3$-Ni &8.5 & 21.7&10.1&0.8 & 0.08&$-$8.4 &$-$2.2&$-$188.08&405.43\\
    \underline{HONH$_3$-Fe} &22.0 &34.0 &13.6&1.6 &0.24 &3.9 & 3.7&7.63&28.34\\
    HONH$_3$-Zn &28.0&33.8&13.0&2.2&0.30&9.1&2.2&7.51&16.72\\
    HONH$_3$-Mg &24.5&41.2&16.9&1.5&0.22&$-$0.7&2.7&8.72&23.35\\
    \hline
    MA-Co(P2$_1$/c)&19.9 &21.1 &8.0&2.5 &0.32 &12.4 &$-$1.8&$-$32.24&59.26\\
    \underline{MA-Mn(P2$_1$/c)}&23.7 &23.4 &8.8 &2.7&0.34 &17.8 &$-$210.5&$-$0.15&32.22\\
    MA-Ni(Pnma)& 27.8&23.1 &8.5 &3.3&0.35 &8.6 &6.3&1.96&22.48\\
    \underline{MA-Zn(P2$_1$/c)} &29.7&27.5&10.2&2.9&0.35&20.9&11.7&4.22&26.36\\
    \hline
    HAZ-Co&30.1&31.5&11.9&2.5&0.33&18.6&12.6&1.77&22.30\\
    HAZ-Mn&24.8&23.1&8.6&2.9&0.35&15.6&$-$20.0&$-$1.73&34.68\\
    HAZ-Zn&28.6&28.8&10.8&2.6&0.33&16.9&32.3&0.82&26.56\\ 
    \underline{HAZ-Mg(Pna2$_1$)}&30.2&31.3&11.8&2.6&0.33&19.6&6.1&3.28&19.88\\
    HAZ-Mg(P2$_1$2$_1$2$_1$)&36.2&34.7&12.9&2.8&0.34&6.6& 2.3&5.28&13.56\\
    \hline
    FA-Mn&22.8&25.4&9.6&2.4&0.32&$-$6.1&53.6&1.07&58.19\\
    \hline
    NH$_4$-Co&33.9&32.0&11.9&2.8&0.34&3.9&$-$9.8&$-$1.73&16.92\\
    NH$_4$-Fe&29.5&27.8&10.3&2.9&0.34&2.5&$-$4.7&$-$4.50&21.31\\
    NH$_4$-Mn&18.5&15.1&5.5&3.4&0.36&0.7&$-$0.9&$-$439.63&381.82\\
    NH$_4$-Zn&34.7&31.5&11.7&3.0&0.35&6.2&$-$8.2&$-$2.04&16.69
    \end{tabular}
    \end{ruledtabular}
    \label{Table:Mech2}
\end{table*}

\begin{figure*}
    \centering
    \includegraphics[width=0.85\textwidth]{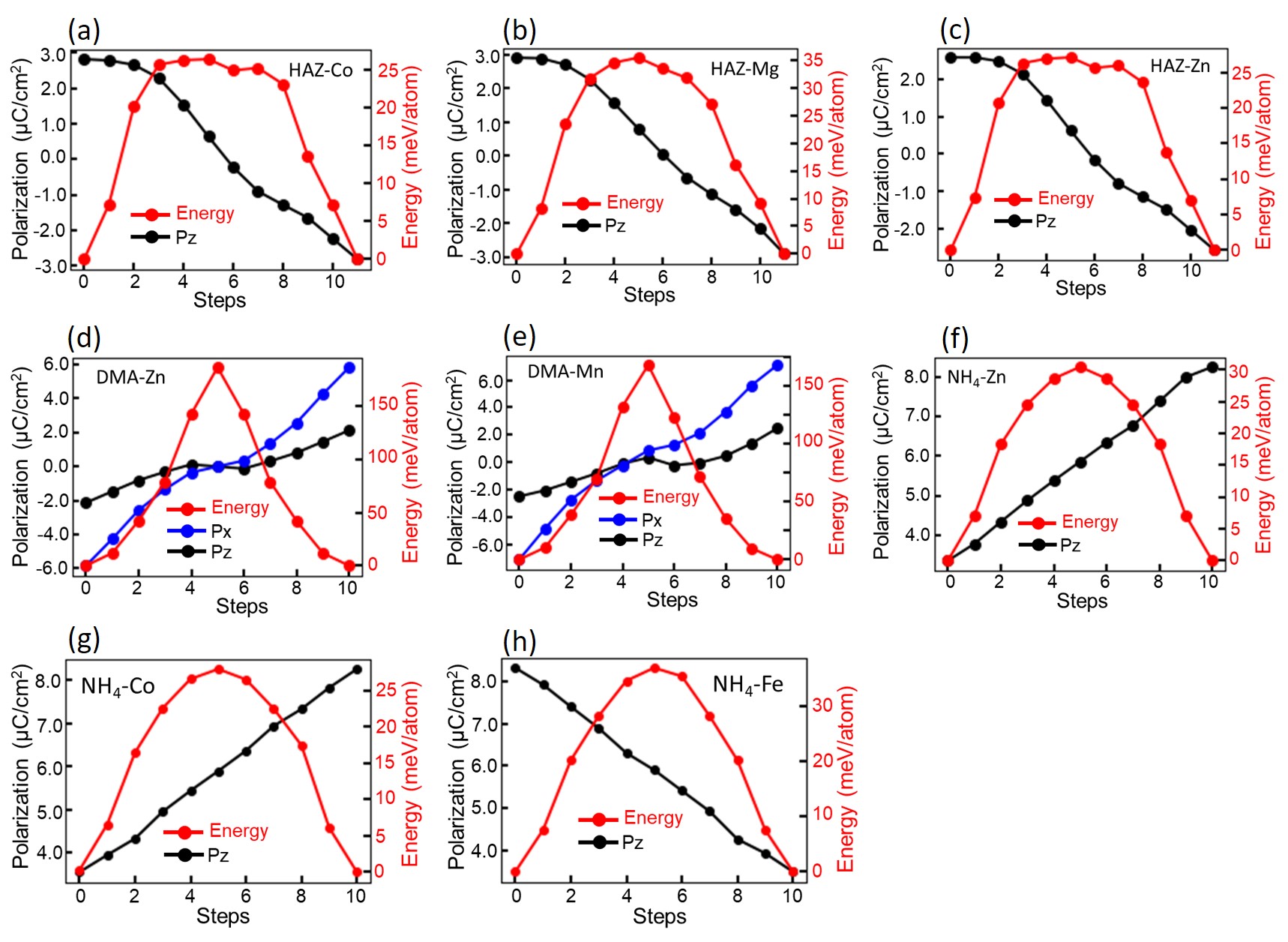}
    \caption{Variation of polarization and energy along the path for  (b) HAZ-Co (c) HAZ-Mg; (d) HAZ-Zn  (e) DMA-Zn (f) DMA-Mn (g) NH$_4$-Zn (h) NH$_4$-Mn (i) NH$_4$-Fe}
    \label{fig_sup1}
\end{figure*}

\begin{figure*}
    \centering
    \includegraphics[width=0.85\textwidth]{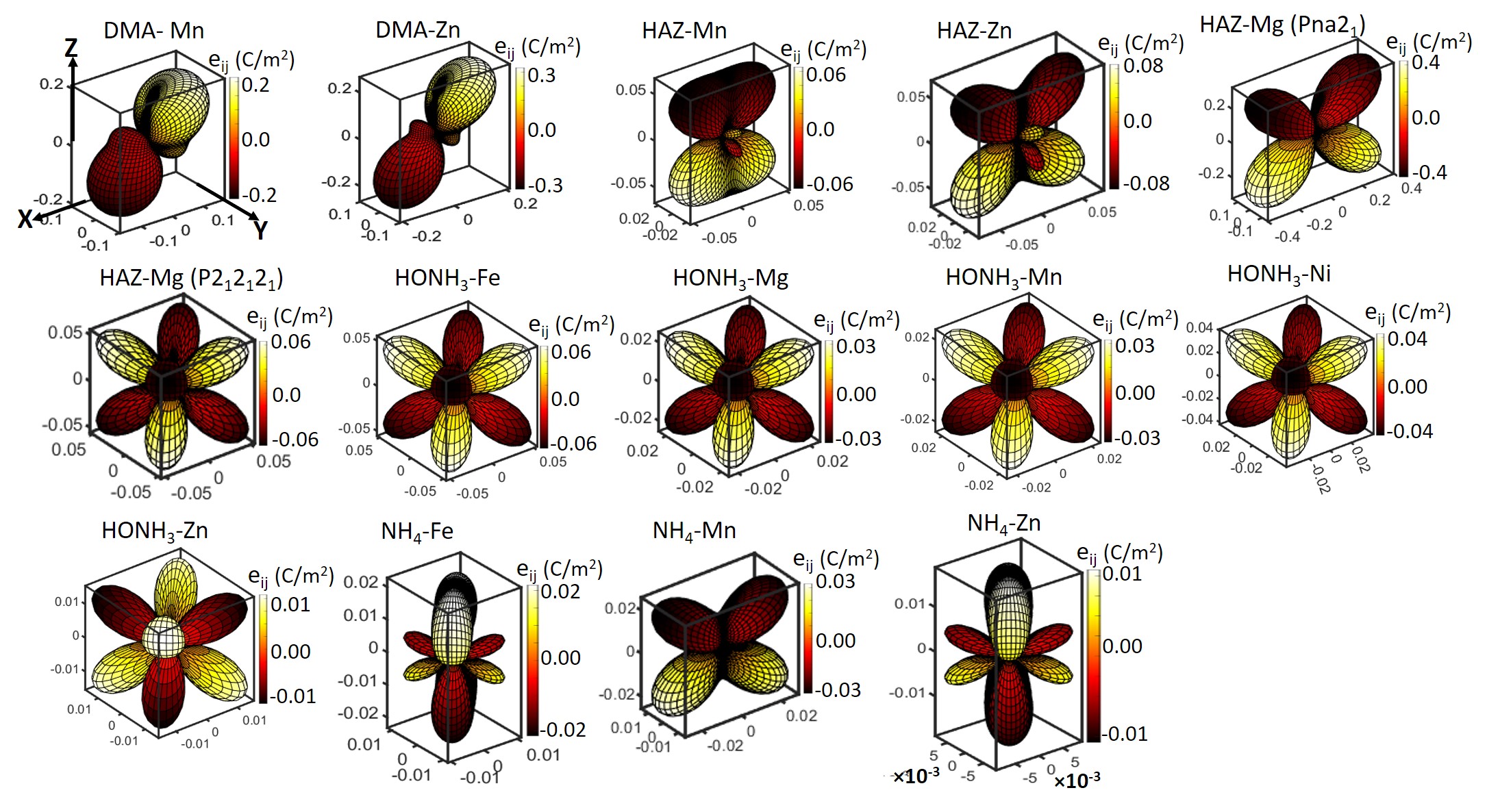}
    \caption{Piezoelectric stress surface for some of the formates studied in this work, as indicated in the titles.}
    \label{fig_sup2}
\end{figure*}

\begin{figure*}
    \centering
    \includegraphics[width=0.85\textwidth]{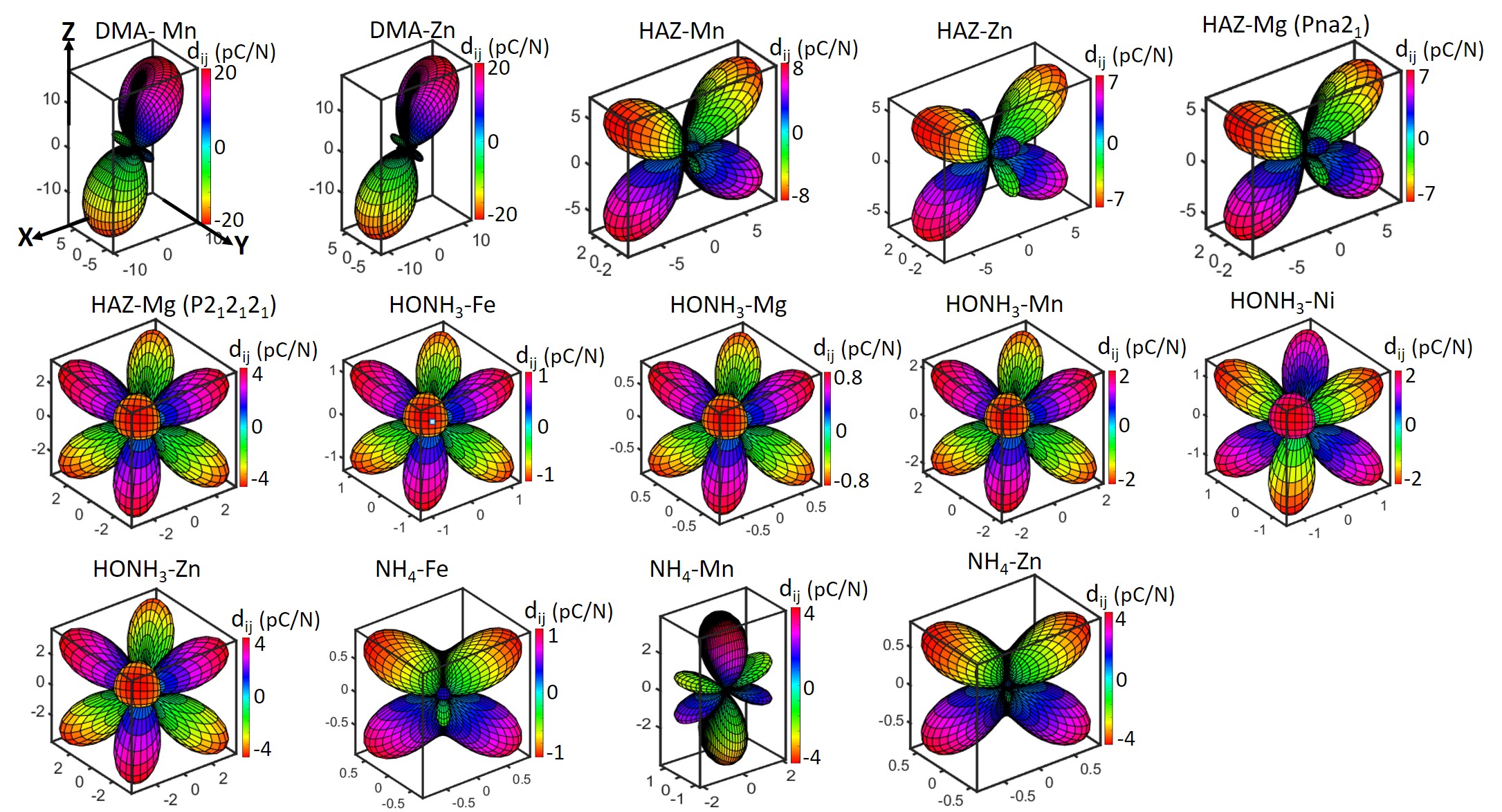}
    \caption{Piezoelectric strain surface for some of the formates studied in this work, as indicated in the titles.}
    \label{fig_sup3}
\end{figure*}

\begin{figure*}
    \centering
    \includegraphics[width=0.85\textwidth]{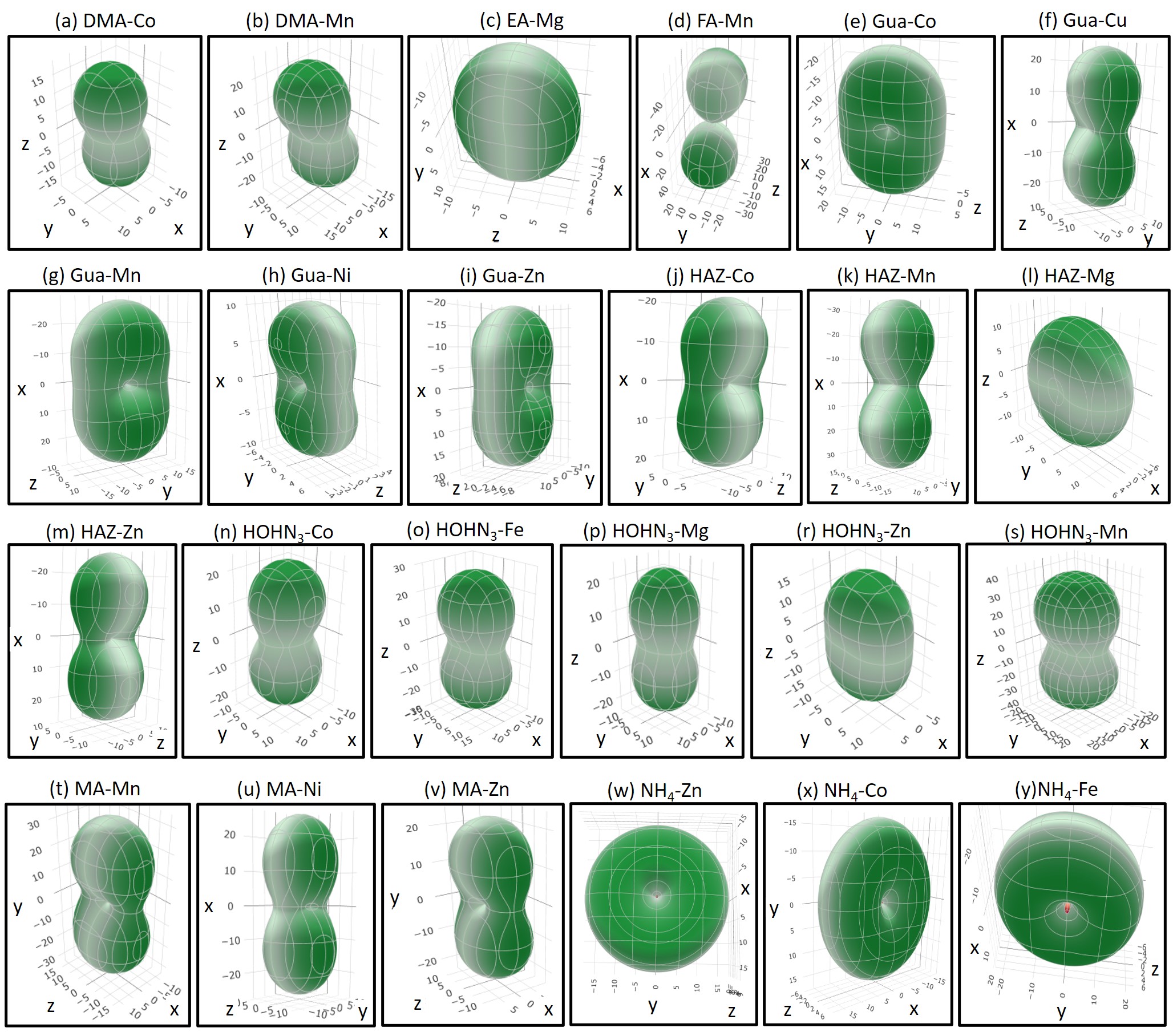}
    \caption{3D plots of linear compressibility for some of the formates studied in this work, as given in the titles. Green and red colors correspond to positive and negative values, respectively.}
    \label{fig_sup4}
\end{figure*}

\end{document}